\documentclass[reprint, aps, pra, footinbib, letterpaper, superscriptaddress]{revtex4-1}

\usepackage[T1]{fontenc} 

\usepackage{amsmath,color}
\usepackage{amssymb}
\usepackage{amsfonts}
\usepackage{amsthm}
\usepackage{mathtools}

\usepackage{algorithm}
\usepackage{algpseudocode}
\usepackage{dsfont}

\usepackage{bbold}
\usepackage{chemformula}
\usepackage{siunitx}

\DeclarePairedDelimiter\ket{\lvert}{\rangle}
\DeclarePairedDelimiterX\braket[2]{\langle}{\rangle}{#1 \delimsize\vert #2}
\DeclarePairedDelimiterX\braket3[3]{\langle}{\rangle}{#1 \delimsize\vert #2 \delimsize\vert #3}

\usepackage{accents}
\newcommand{\dbtilde}[1]{\accentset{\approx}{#1}}

\newcommand{\ve}{\mathbf{e}}
\newcommand{\vu}{\mathbf{u}}

\newcommand{\vmu}{\boldsymbol{\mu}}

\newcommand{\vR}{\mathbf{r}}

\newcommand{\vk}{\mathbf{k}}

\newcommand{\vxi}{\boldsymbol{\xi}}

\newcommand{\hH}{\hat{H}}

\newcommand{\avg}[1]{\left\langle #1\right\rangle}

\begin{document}

	\title{Cavity Molecular Dynamics Simulations of Liquid Water under Vibrational  Ultrastrong Coupling}
	
	\author{Tao E. Li}%
	\affiliation{Department of Chemistry, University of Pennsylvania, Philadelphia, Pennsylvania 19104, USA}
	
	\author{Joseph E. Subotnik}
	\affiliation{Department of Chemistry, University of Pennsylvania, Philadelphia, Pennsylvania 19104, USA}

	\author{Abraham Nitzan} 
	\email{anitzan@sas.upenn.edu}
	\affiliation{Department of Chemistry, University of Pennsylvania, Philadelphia, Pennsylvania 19104, USA}
	\affiliation{School of Chemistry, Tel Aviv University, Tel Aviv 69978, Israel}

	\begin{abstract}
		We simulate vibrational strong (VSC) and ultrastrong coupling (V-USC) for liquid water with classical molecular dynamics simulations. When the cavity modes are resonantly coupled to the \ch{O-H} stretch mode of liquid water, the infrared spectrum shows asymmetric Rabi splitting. The lower polariton (LP) may be suppressed or enhanced relative to the upper polariton  (UP) depending on the frequency of the cavity mode. Moreover, although the static properties and the translational diffusion of water are not changed under VSC or V-USC, we do find the modification of the orientational autocorrelation function of \ch{H2O} molecules especially under V-USC, which could play a role in ground-state chemistry. 
	\end{abstract}

	\maketitle
	
	\section{Introduction}
	Strong light-matter interactions between a vibrational mode of molecules and a cavity mode have attracted great attention of late  \cite{Herrera2019}. The signature of strong interactions is the formation of lower (LP) and upper (UP) polaritons, which are manifested in the Rabi splitting of a vibrational peak in the molecular infrared (IR) spectrum. According to the normalized ratio ($\eta$) between the Rabi splitting frequency  ($\Omega_N$) and the original vibrational frequency ($\omega_0$), or $\eta = \Omega_N/ 2\omega_0$,  one often classifies $0 < \eta < 0.1$ as vibrational strong coupling (VSC) and  $\eta > 0.1$ as vibrational ultrastrong coupling (V-USC) \cite{FriskKockum2019}. The investigation of VSC or V-USC in liquid phase was initially suggested by Ebbesen \textit{et al} \cite{Shalabney2015,George2015,George2016}, and it was later  found experimentally that VSC or V-USC can modify the ground-state chemical reaction rates of molecules even without external pumping \cite{Thomas2016}. This exotic catalytic effect provides a brand new way to control chemical reactions remotely. As such, there has been a recent push to understand the origins and implications of VSC and V-USC.
	
	While the experimental side has focused on the search for large catalytic effects \cite{Lather2019,Hiura2018,Thomas2019_science,Vergauwe2019} as well as understanding polariton relaxation dynamics through two-dimensional IR (2D-IR) spectroscopy \cite{Xiang2018,Xiang2019}, on the theoretical side, the nature of VSC and V-USC remains obscured. On the one hand,  Rabi splitting can be easily modeled by, e.g., diagonalizing a model Hamiltonian in the singly excited manifold \cite{Hopfield1958,Hernandez2019,Du2018} or solving equations of motion classically for a set of one-dimensional (1D) harmonic oscillators \cite{Rudin1999,F.Ribeiro2018}. On the other hand, a robust explanation of the catalytic effect of VSC or V-USC remains illusive \cite{Galego2019,Campos-Gonzalez-Angulo2019,Hiura2019,Li2020Origin}. For example, as recently shown by us and others \cite{Li2020Origin,Campos-Gonzalez-Angulo2020,Zhdanov2020}, the classical potential of mean force along a reaction pathway is not changed by usual VSC or V-USC setups for standard experiments of interest. Moreover, as demonstrated below, any static equilibrium property of a molecule is not changed under VSC or V-USC when nuclei and photons are treated classically. These findings, unfortunately, show that one cannot explain the observed effect under VSC or V-USC from a static and classical view of point.  From such a conclusion, one possible hypothesis of the manifestations of VSC or V-USC effect on chemical rates should  arise from the modification of non-equilibrium, or dynamical, properties of molecules under VSC or V-USC.

	The first step towards proving the above hypothesis is to ascertain whether or not any dynamical property of molecules is actually changed for a realistic experiment, a goal which forms the central objective of this manuscript. 
	In order to investigate whether such modification occurs, below we will model VSC and V-USC using cavity molecular dynamics (MD) simulation, where the nuclei are evolved under a realistic electronic ground-state potential surface.
	Such an approach is an extension of the usual simplified 1D models where the matter side is evolved as two-level systems \cite{Goto2005,Li2020Quasi,Hoffmann2019Benchmark} or coupled harmonic oscillators \cite{Rudin1999,Santhosh2016,Sukharev2018,F.Ribeiro2018}. Although such simplified models are adequate enough for studying Rabi splitting qualitatively by fitting experimental parameters, these models usually ignore  translation, rotation, collision, as well as the intricate structure of molecular motion, all of which are crucial for determining the dynamic properties of molecules. Therefore, explicit cavity MD simulations become a more appropriate approach for studying all dynamic properties. 
	Moreover, even though one can find a Rabi splitting from 1D models, performing cavity MD simulations is also very helpful for as providing more details about the IR spectrum and this approach can be used to benchmark the validity of 1D models under various conditions.

	There have been a few flavors of cavity MD schemes for \textit{electronic} strong coupling \cite{Flick2017,Luk2017,Groenhof2019}. For example, Luk \textit{et al} applied multiscale quantum mechanics/molecular mechanics (QM/MM) simulation for studying the dynamics of electronic polaritons for Rhodamine molecules \cite{Luk2017}.
	By contrast,
	MD simulations for \textit{vibrational} strong coupling (VSC and V-USC), to our best knowledge, have not been extensively studied before. Therefore, below we will first establish a framework for cavity MD simulation including implementation details, and second we will investigate the Rabi splitting and the dynamical properties of liquid water.

	The motivation for studying liquid water is two-fold: (i) Among common liquids, water shows strong Rabi splitting and strong catalytic effects under VSC or V-USC \cite{Vergauwe2019,Hiura2019water,Hiura2018}. More interesting, when the cavity mode is resonantly coupled to the \ch{O-H} stretch mode, experiments \cite{Vergauwe2019} have observed that the intensity of the vibrational LP peak is much smaller than the UP peak in the IR spectrum, an observation that cannot be accounted for by standard strong coupling models. (ii) MD simulations of water outside the cavity have been extensively studied and  good agreement with experiments can be achieved \cite{Abascal2005,Habershon2009,Corcelli2004}. 
	Extending such simulations to include coupling to cavity modes is expected to show the cavity-induced spectral changes and provides numbers that are directly comparable to experimental results.
	
	
	\section{General Theory of V-USC}\label{sec:theory}
	The full-quantum Hamiltonian for light-matter interactions reads \cite{Cohen-Tannoudji1997,Li2020Origin}:
	\begin{subequations}\label{eq:H_qed}
		\begin{align}
		\hat{H}_{\text{QED}} = \hat{H}_{\text{M}} + 
		\hat{H}_{\text{F}}
		\end{align}
	Here, $\hat{H}_{\text{M}}$ denotes the conventional (kinetic + potential) Hamiltonian for the molecular system
		\begin{align}
			\hat{H}_{\text{M}} = \sum_{i} \frac{\hat{\mathbf{p}}_i^2}{2 m_i} +  \hat{V}_{\text{Coul}}\left(\{\hat{\vR}_i\}\right) 
		\end{align}
	 where $m_i$, $\hat{\mathbf{p}}_i$, $\hat{\mathbf{r}}_i$ denote the mass, momentum operator, and position operator for the $i$-th particle (nucleus or electron), respectively, and $\hat{V}_{\text{Coul}}\left(\{\hat{\vR}_i\}\right)$ denotes the Coulombic interaction operator between all nuclei and electrons. 
	 	Under the long-wave approximation, the field-related Hamiltonian $\hat{H}_{\text{F}}$ reads
	 \begin{align}\label{eq:H_qed-3}
	 	\hat{H}_{\text{F}} = \sum_{k,\lambda}
	 	\frac{1}{2}\omega_{k,\lambda}^2 \hat{q}_{k,\lambda}^2 + \frac{1}{2}\left(\hat{p}_{k, \lambda} - \frac{1}{\sqrt{\Omega\epsilon_0}} \hat{\vmu}_S\cdot \vxi_{\lambda}  \right)^2
	 \end{align}
	\end{subequations}
	  where $\omega_{k,\lambda}$, $\hat{q}_{k, \lambda}$, $\hat{p}_{k,\lambda}$  denote the frequency, position operator, and momentum operator for a photon with wave vector $\mathbf{k}$ and polarization direction $\vxi_{\lambda}$, and the index $\lambda = 1,2$ denotes the two polarization directions which satisfy $\mathbf{k}\cdot \vxi_{\lambda} = 0$. In free space, the dispersion relation gives $\omega_{k,\lambda} = c |\mathbf{k}| = ck$. $\epsilon_0$ and $\Omega$ denote the vacuum permittivity and the cavity volume.  $\hat{\vmu}_S$ denotes the dipole operator for the whole molecular system:
	$\hat{\vmu}_S = \sum_{i} Z_i e \hat{\vR}_i$,
	where $e$ denotes the electron charge and $Z_i e$ denotes the charge for the $i$-th particle (nucleus or electron). $\hat{\vmu}_S$ can also be grouped into a summation of molecular dipole moments (indexed by $n$):
	$\hat{\vmu}_S = \sum_{n=1}^{N} \hat{\vmu}_n;  \ \ \   \hat{\vmu}_n = \sum_{j \in n} Z_j e \hat{\vR}_j$.
	Note that the self-dipole term in Eq. \eqref{eq:H_qed-3} (i.e., the $\hat{\vmu}_S^2$ term in the expanded square) is of vital importance in describing USC 
	and is needed to render the nuclear motion stable; see Refs. \cite{Rokaj2018,Schafer2020,Hoffmann2020} for details. Because we will not neglect $\hat{\vmu}_S^2$ below, our simulation is valid for both VSC and U-VSC.
	
	When the cavity mode frequency is within the timescale of the nuclear dynamics, the Born-Oppenheimer approximation implies that electrons stay in the  ground state. Therefore, we will project the quantum Hamiltonian \eqref{eq:H_qed} onto the electronic ground state, $\hH_\text{QED}^{\text{G}} = \braket3{\Psi_\text{G}}{\hH_{\text{QED}}}{\Psi_\text{G}}$, where $\ket{\Psi_\text{G}}$ denotes the electronic ground state for the whole molecular system. Furthermore, under the Hartree approximation, $\ket{\Psi_\text{G}}$ can be approximated as a product of the electronic ground states for individual molecules: $\ket{\Psi_\text{G}} = \prod_{n=1}^{N} \ket{\psi_{ng}}$. After such a projection on the electronic ground state, the Hamiltonian \eqref{eq:H_qed} reduces to
	\begin{subequations}\label{eq:H_QED_G}
		\begin{equation}
		\begin{aligned}
		\hH_\text{QED}^{\text{G}} = \ & \hH_{\text{M}}^{\text{G}} 
		+ \hH_{\text{F}}^{\text{G}} .
		\end{aligned}
		\end{equation}
		Here, the ground-state molecular Hamiltonian $\hH_{\text{M}}^{\text{G}} = \braket3{\Psi_\text{G}}{\hH_{\text{M}}}{\Psi_\text{G}}$ depends on the nuclear degrees of freedom only, and can be expressed as
		\begin{align}\label{eq:H_QED_G-2}
		\hH_{\text{M}}^{\text{G}} = \sum_{n=1}^{N}\left(\sum_{j\in n} \frac{\hat{\mathbf{P}}_{nj}^2}{2 M_{nj}} + \hat{V}^{(n)}_{g}(\{\hat{\mathbf{R}}_{nj}\})\right) + 
		\sum_{n=1}^{N}\sum_{l>n}
		\hat{V}_{\text{inter}}^{(nl)}
		\end{align}
		where the capital letters $\hat{\mathbf{P}}_{nj}$, $\hat{\mathbf{R}}_{nj}$, and $M_{nj}$  denote the momentum operator, position operation, and mass for the $j$-th nuclus in molecule $n$, $\hat{V}^{(n)}_{g}$ denotes the intramolecular potential for molecule $n$, and $\hat{V}_{\text{inter}}^{(nl)}$ denotes the intermolecular interactions between molecule $n$ and $l$.
		The field-related Hamiltonian becomes \cite{Li2020Origin}
		\begin{equation}\label{eq:HF_G}
		\begin{aligned}
		\hH_{\text{F}}^{\text{G}}  = \ & \sum_{k,\lambda}
		\frac{1}{2}\omega_{k,\lambda}^2 \hat{q}_{k,\lambda}^2 + \frac{1}{2}\left(\hat{p}_{k, \lambda} - \sum_{n=1}^{N} \frac{1}{\sqrt{\Omega\epsilon_0}} \hat{d}_{ng, \lambda}  \right)^2 \\
		& +
		\sum_{k,\lambda}\sum_{n=1}^{N} \frac{1}{2\Omega \epsilon_0} \braket3{\psi_{ng}}{\delta \hat{d}_{ng, \lambda}^2}{\psi_{ng}} 
		\end{aligned}
		\end{equation}
	\end{subequations}
	where we define $\hat{d}_{ng, \lambda} \equiv \braket3{\psi_{ng}}{\hat{\vmu}_{n} }{\psi_{ng}} \cdot \vxi_{\lambda}$ and $\delta \hat{d}_{ng, \lambda} \equiv \hat{\vmu}_{n} \cdot \vxi_{\lambda} - \hat{d}_{ng, \lambda}$.
	Note that, since Coulombic interactions are modified by proximity to dielectric boundaries, in the cavity, the intermolecular interactions $
	\hat{V}_{\text{inter}}^{(nl)}$ 
	in Eq. \eqref{eq:H_QED_G-2}
	may differ from the free-space form \cite{Takae2013,DeBernardis2018}. However, as we have argued before  \cite{Li2020Origin}, for standard VSC setups with a cavity length on the order of microns,  $\hat{V}_{\text{inter}}^{(nl)}$ should be nearly identical to those in free space \cite{footnote_joe}. Similarly, on the last line in Eq. \eqref{eq:HF_G}, the self-dipole fluctuation term $\frac{1}{2\Omega \epsilon_0} \braket3{\psi_{ng}}{\delta \hat{d}_{ng, \lambda}^2}{\psi_{ng}} $, which denotes the cavity modification of the single-molecule potential, should also be very small for standard VSC setups where micron-length cavities are used. Therefore, in what follows, we will 
	assume that $\hat{V}_{\text{inter}}^{(nl)}$ take the free-space form and also
	neglect the self-dipole fluctuation term. However, we emphasize that, for smaller cavities, both the change of intermolecular interactions and the self-dipole fluctuation may play an important role in  ground-state chemistry as already discussed in different contexts \cite{Flick2017,Galego2019,Schafer2020}, a fact which needs further investigation.
	
	In MD simulations, a standard potential is a function of positions only. In Eq. \eqref{eq:HF_G}, however, the momenta of photons are coupled directly to the molecular dipole moments (which are a function of the nuclear positions of the molecules). 
	However, since photons are harmonic oscillators, we may exchange the momentum and position of each photon, so that  Eq. \eqref{eq:HF_G} can be rewritten as
	\begin{equation}\label{eq:H_QM}
	\begin{aligned}
	\hH_{\text{F}}^{\text{G}} & = \sum_{k,\lambda}
	\frac{\hat{\widetilde{p}}_{k, \lambda}^2}{2 m_{k, \lambda}}  \\
	&+ 
	\frac{1}{2} m_{k,\lambda}\omega_{k,\lambda}^2 \Bigg (
	\hat{\widetilde{q}}_{k,\lambda}  + \sum_{n=1}^{N} \frac{\hat{d}_{ng, \lambda}}{\omega_{k,\lambda}\sqrt{\Omega\epsilon_0 m_{k,\lambda}}} 
	\Bigg )^2 
	\end{aligned}
	\end{equation}
	Here, to be compatible with standard MD simulations (which requires the information of mass for particles), an auxiliary mass $m_{k, \lambda}$ for each photon is also introduced: $\hat{p}_{k, \lambda} = \hat{\widetilde{p}}_{k, \lambda}/\sqrt{m_{k, \lambda}}$ and $\hat{q}_{k, \lambda} = \sqrt{m_{k, \lambda}} \hat{\widetilde{q}}_{k, \lambda}$.
	Note that the auxiliary mass of photon does not alter any dynamics and serves only as a convenient notation for further MD treatment.
	
	\section{Classical Molecular Dynamics}
	The quantum Hamiltonian in Eq. \eqref{eq:H_QM}, although depending only on the nuclear and photonic degrees of freedom, is still too expensive to evolve exactly. The simplest approximation we can make is the classical approximation, i.e., all quantum operators are mapped to the corresponding classical observables. After applying the periodic boundary condition for the molecules,
	the equations of motion for the coupled nuclei-photonic system become (see SI Appendix, Sec. 1):
	\begin{widetext}
	\begin{subequations}\label{eq:EOM_MD_PBC}
		\begin{align}
		M_{nj}\ddot{\mathbf{R}}_{nj} &= \mathbf{F}_{nj}^{(0)}  - 
		\sum_{k,\lambda}
		\left(\widetilde{\varepsilon}_{k,\lambda} \dbtilde{q}_{k,\lambda}
		+ \frac{\widetilde{\varepsilon}_{k,\lambda}^2}{m_{k,\lambda} \omega_{k,\lambda}^2} \sum_{l=1}^{N_{\text{sub}}} d_{lg,\lambda}
		\right) 
		\frac{\partial d_{ng, \lambda}}{\partial \mathbf{R}_{nj}}
		\\
		m_{k,\lambda}\ddot{\dbtilde{q}}_{k,\lambda} &= - m_{k,\lambda}\omega_{k,\lambda}^2 \dbtilde{q}_{k,\lambda}
		-\widetilde{\varepsilon}_{k,\lambda} \sum_{n=1}^{N_{\text{sub}}}d_{ng,\lambda}
		\end{align}
	\end{subequations}
	\end{widetext}
	Here,  $\mathbf{F}_{nj}^{(0)} = - \partial V_g^{(n)} / \partial \mathbf{R}_{nj} - \sum_{l\neq n} \partial V_{\text{inter}}^{(nl)}/\partial \mathbf{R}_{nj}$ denotes the cavity-free force on each nuclei. We have defined $\dbtilde{q}_{k,\lambda} = \widetilde{q}_{k,\lambda} / \sqrt{N_{\text{cell}}}$, and the effective coupling strength $\widetilde{\varepsilon}_{k,\lambda} = \sqrt{N_{\text{cell}} m_{k,\lambda} \omega_{k,\lambda}^2/\Omega \epsilon_0}$, where $N_{\text{cell}}$ denotes the number of the periodic simulation cells for \ch{H2O} molecules. $N_{\text{sub}}$ denotes the number of molecules in a single simulation cell and the total number of molecules is $N=N_{\text{sub}}N_{\text{cell}}$. More details on implementation and simulations are explained in Materials and Methods and SI Appendix.

	\section{Results}\label{sec:results}
	
	\subsection{Asymmetric Rabi Splitting}
	
	\begin{figure}
		\centering
		\includegraphics[width=0.7\linewidth]{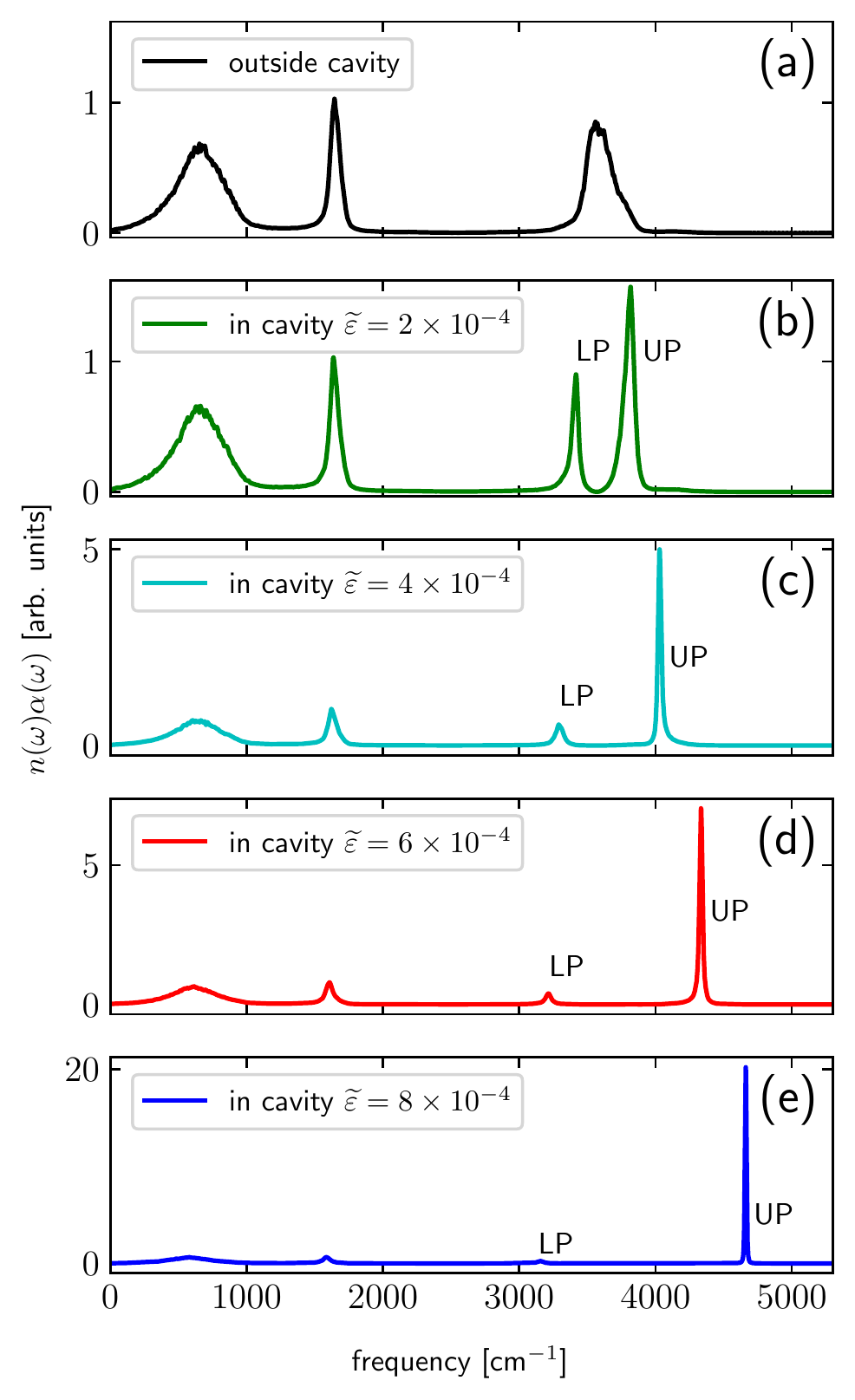}
		\caption{Simulated IR spectrum of liquid water under VSC or V-USC. From top to bottom, we plot the results (a) outside the cavity, or inside the cavity with effective coupling strength $\widetilde{\varepsilon}$ as (b) $2\times 10^{-4}$, (c) $4\times 10^{-4}$, (d) $6\times 10^{-4}$, and (e) $8\times 10^{-4}$ a.u.. All other simulation details are listed in Materials and Methods. Note that, as  $\widetilde{\varepsilon}$ increases, the LP peak is suppressed and the UP peak is enhanced.}
		\label{fig:IR}
	\end{figure}

	The signature of VSC is the collective Rabi splitting in the IR spectrum. 	In our MD simulations, the IR spectrum is calculated by linear response theory. For isotropic liquids, the absorption coefficient $\alpha(\omega)$ is expressed as the Fourier transform of the autocorrelation function of the total dipole moment $\vmu_S$  \cite{McQuarrie1976,Gaigeot2003,Habershon2008,Nitzan2006,footnote_abe}:
	\begin{equation}\label{eq:IR_def}
	n(\omega)\alpha(\omega) = \frac{\pi \beta \omega^2 }{3 \epsilon_0 V c } \frac{1}{2\pi} \int_{-\infty}^{+\infty} dt \ e^{-i\omega t} \avg{\vmu_S(0) \vmu_S(t)} 
	\end{equation}
	Here, $n(\omega)$ denotes the refractive index and $V$ denotes the volume of the system (i.e., the simulation cell). The factor $\omega^2$ arises from the absorbed photon energy by the liquid. See SI Appendix Sec. 1 for  calculating $\vmu_S$.
	For VSC and V-USC experiments, however, because the experimental setups usually detect an IR spectrum by sending light along the cavity direction (which means the $\vk$ direction of light is along the $z$-axis)  \cite{Hiura2019water},  we need to modify the above equation to
	\begin{equation}\label{eq:IR_equation_cavity}
	\begin{aligned}
	n(\omega)\alpha(\omega) &= \frac{\pi \beta \omega^2}{2\epsilon_0 V c} \frac{1}{2\pi}  \int_{-\infty}^{+\infty} dt \ e^{-i\omega t}  \\
	&\times \avg{\sum_{i=x, y}\left(\vmu_S(0)\cdot \ve_i\right) \left(\vmu_S(t)\cdot \ve_i\right)} 
	\end{aligned}
	\end{equation}
	where $\ve_i$ denotes the unit vector along direction $i=x, y$. Eq. \eqref{eq:IR_equation_cavity} states that the average is performed only along the polarization directions of the detecting signal (i.e., the $x$ and $y$ directions here). When the incident light is unpolarized these two directions are of course equivalent.
	
	Fig. \ref{fig:IR}a plots the simulated IR spectrum of liquid water outside the cavity. The \ch{O-H} stretch peaks around $\sim 3550 \text{\ cm}^{-1}$, which is slightly different from experiment ($\sim 3400 \text{\ cm}^{-1}$). Noted that a more accurate \ch{O-H} stretch peak can be simulated by performing path-integral calculations  instead of a classical simulation \cite{Habershon2009}.
	
	For the case that the frequency of the two photon modes (with polarization directions perpendicular to the cavity direction) are both set to be at resonance with the \ch{O-H} stretch ($3550 \text{\ cm}^{-1}$), Figs. \ref{fig:IR}(b)-(d) plot the simulated IR spectrum; the effective coupling strength $\widetilde{\varepsilon}$ is set as $2\times 10^{-4}$, $4\times 10^{-4}$,   $6\times 10^{-4}$, and $8\times 10^{-4}$ a.u., respectively. Clearly, when the cavity modes are coupled to the \ch{H2O} molecules,  the broad \ch{O-H} stretch peak is spit into a pair of narrower LP and UP peaks.
	This result agrees with the previous theoretical and experimental work that the inhomogeneous broadening of the vibrational peak does not lead to the broadening of the polariton peaks \cite{Houdre1996,Long2015}.
	More interestingly, our simulation results also suggest that the UP and LP peaks can be largely asymmetric especially when $\widetilde{\varepsilon}$ is large, which agrees with experimental findings at least qualitatively \cite{Vergauwe2019}.
	
		\begin{figure}
			\centering
		\includegraphics[width=0.7\linewidth]{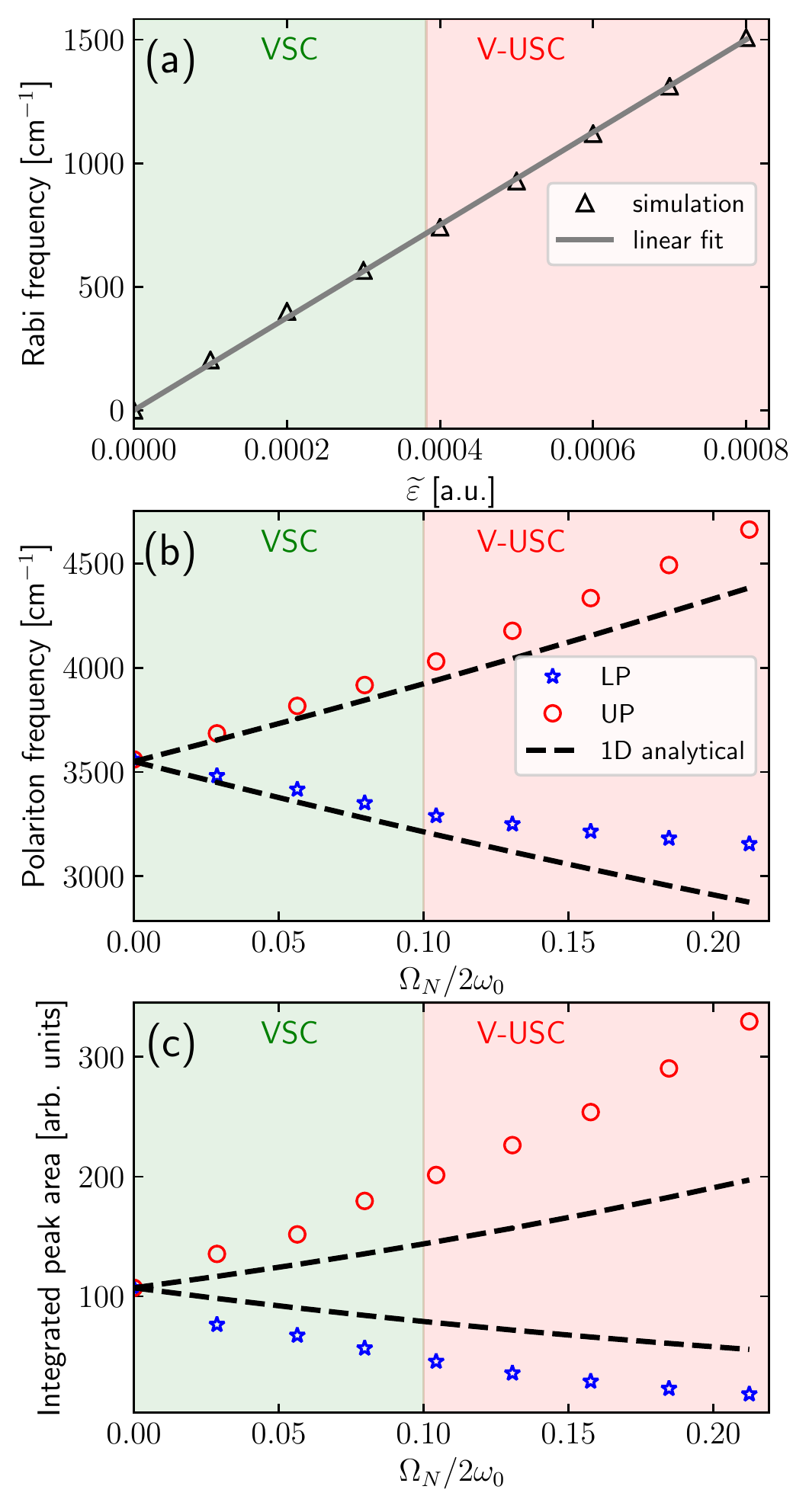}
		\caption{(a) Rabi frequency ($\Omega_N$) as a function of the effective coupling strength $\widetilde{\varepsilon}$ for liquid water. (b) Polariton frequency and (c) integrated peak area of polaritons as a function of normalized Rabi frequency ($\Omega_N/2\omega_0$).  All simulation details are the same as Fig. \ref{fig:IR}. In Fig. \ref{fig:Rabi_splitting}a the simulation data (black triangles) are fit linearly (gray line). In Fig. \ref{fig:Rabi_splitting}b-c the simulation data (blue stars for LP and red circles for UP) are compared with the analytical expressions from a simplified 1D model (Eqs. \eqref{eq:omega_LP_UP} and \eqref{eq:I_area_LP_UP}), where the parameters are given as $\omega_0 = \omega_c = 3550 \text{\ cm}^{-1}$ and $\Omega_N$ is taken as the values in Fig. \ref{fig:Rabi_splitting}a.}
		\label{fig:Rabi_splitting}
	\end{figure}

	In Fig. \ref{fig:Rabi_splitting}a we plot the Rabi splitting frequency (the difference between the UP and LP frequencies, or $\omega_{+} - \omega_{-}$) as a function of $\widetilde{\varepsilon}$. The simulation data (black triangles) can be  fit with a linear ansatz (gray line) very well. As mentioned above, because  $\widetilde{\varepsilon} = \sqrt{N_{\text{cell}}} \varepsilon \propto \sqrt{N}$, Fig. \ref{fig:Rabi_splitting}a demonstrates that the Rabi splitting is proportional to the square root of the total number of molecules, which agrees with 
	theoretical expectation and experimental observation\cite{Meystre2007,Hiura2019water}:
	\begin{equation}
	\omega_{+} - \omega_{-} = \Omega_N \equiv 2g_0 \sqrt{N} 
	\end{equation}
	where $g_0$ denotes the coupling constant between a single molecule and the photon mode.

	Of particular interest is the asymmetric nature of the LP and UP: this asymmetry is manifest in two aspects. As shown in Fig. \ref{fig:Rabi_splitting}b-c,
	both the polariton frequencies and the integrated peak areas of the LP (blue stars) and UP (red circles) show asymmetric scalings as a function of the normalized Rabi frequency ($\Omega_N/2\omega_0$, where $\Omega_N$ is taken from Fig. \ref{fig:Rabi_splitting}a), especially in the V-USC limit (the red-shadowed region). 	
	Note that the standard treatment of collective Rabi splitting does not account for this asymmetry and the observation of the suppression (or enhancement) of the LP (or the UP) in Ref. \cite{Vergauwe2019} was explained by the higher absorption of water and gold cavity mirrors in the LP region.
	Some insight into the origin of this asymmetry can be obtained from a simple 1D model where $N$ independent harmonic oscillators interact with a single photon mode. By taking the self-dipole term into account (to describe V-USC), we obtain  (see SI Appendix, Sec. 2)
	\begin{subequations}\label{eq:omega_LP_UP}
	\begin{align}\label{eq:omega_LP_UP-1}
	\omega_{\pm}^2 &= \frac{1}{2}\left[\omega_0^2 + \Omega_N^2 + \omega_c^2 \pm \sqrt{(\omega_0^2 + \Omega_N^2 + \omega_c^2)^2 - 4\omega_0^2\omega_c^2}\right] \\
	\label{eq:omega_LP_UP-2}
	& = \omega_0^2 + \frac{\Omega_N^2}{2} \pm \Omega_N\sqrt{\omega_0^2 + \frac{\Omega_N^2}{4}} \ \ \ (\text{when\ } \omega_c = \omega_0)
	\end{align}
	\end{subequations}
	where $\omega_0$ and $\omega_c$ denote the frequencies of the harmonic oscillators and the photon mode. 
	Given $\omega_0 = \omega_c = 3550 \text{\ cm}^{-1}$ and $\Omega_N$ in Fig. \ref{fig:Rabi_splitting}a, we have plotted Eq. \eqref{eq:omega_LP_UP} (the black dashed lines) in Fig. \ref{fig:Rabi_splitting}b. 
	We see that this analytical result already shows some asymmetry in the positions of the polariton peaks when plotted versus $\Omega_N$.
	While  Eq. \eqref{eq:omega_LP_UP} agrees with our simulation data very well in the VSC limit (the green-shadowed region), the simulation data seem to be more asymmetric than Eq. \eqref{eq:omega_LP_UP} in the V-USC limit. Such disagreement may arise from the strong intermolecular interactions between \ch{H2O} molecules, which is completely ignored in the simplified 1D model of the SI Appendix.
	
	Likewise, the simplified 1D model in the SI Appendix also suggests that the integrated peak areas of the LP and UP are
	\begin{subequations}\label{eq:I_area_LP_UP}
		\begin{align}
		I_{\text{LP}} &\propto \omega_{-}^2 \sin^2\left(\frac{\theta}{2}\right) \\
		I_{\text{UP}} &\propto \omega_{+}^2 \cos^2\left(\frac{\theta}{2}\right)
		\end{align}
	\end{subequations}
	where $\tan\left(\theta\right) = 2\omega_c \Omega_N / \left(\omega_0^2 + \Omega_N^2 - \omega_c^2\right)$. 	Again, as shown in Fig. \ref{fig:Rabi_splitting}c, Eq. \eqref{eq:I_area_LP_UP} (black dashed lines) matches the simulation data roughly but not quantitatively, which may come from ignoring all the  intermolecular interactions in the 1D model.
	Nevertheless, from Eq. \eqref{eq:I_area_LP_UP}, we  find that the asymmetry in the IR spectrum comes from two factors: (i) the factor $\omega_{\pm}^2$ and (ii) the angular part  $\sin^2\left(\frac{\theta}{2}\right)$ or $\cos^2\left(\frac{\theta}{2}\right)$. While the first part originates from the  absorbed photon energies associated with the vibration modes and is  universal for all IR spectrum (so that it is trivial), the second factor is quite nontrivial: at resonance ($\omega_0 = \omega_c$) one would naively assume that $\sin^2\left(\frac{\theta}{2}\right) = \cos^2\left(\frac{\theta}{2}\right)$ and this is true if one ignores the self-dipole term (which means ignoring the $\Omega_N^2$ term  in $\tan\left(\theta\right)$; see SI Appendix for details). However, when the self-dipole term is considered, one finds $\sin^2\left(\frac{\theta}{2}\right) < \cos^2\left(\frac{\theta}{2}\right)$, which leads to an additional suppression of the LP and the enhancement of the UP.
	
	\begin{figure}
		\centering
		\includegraphics[width=0.7\linewidth]{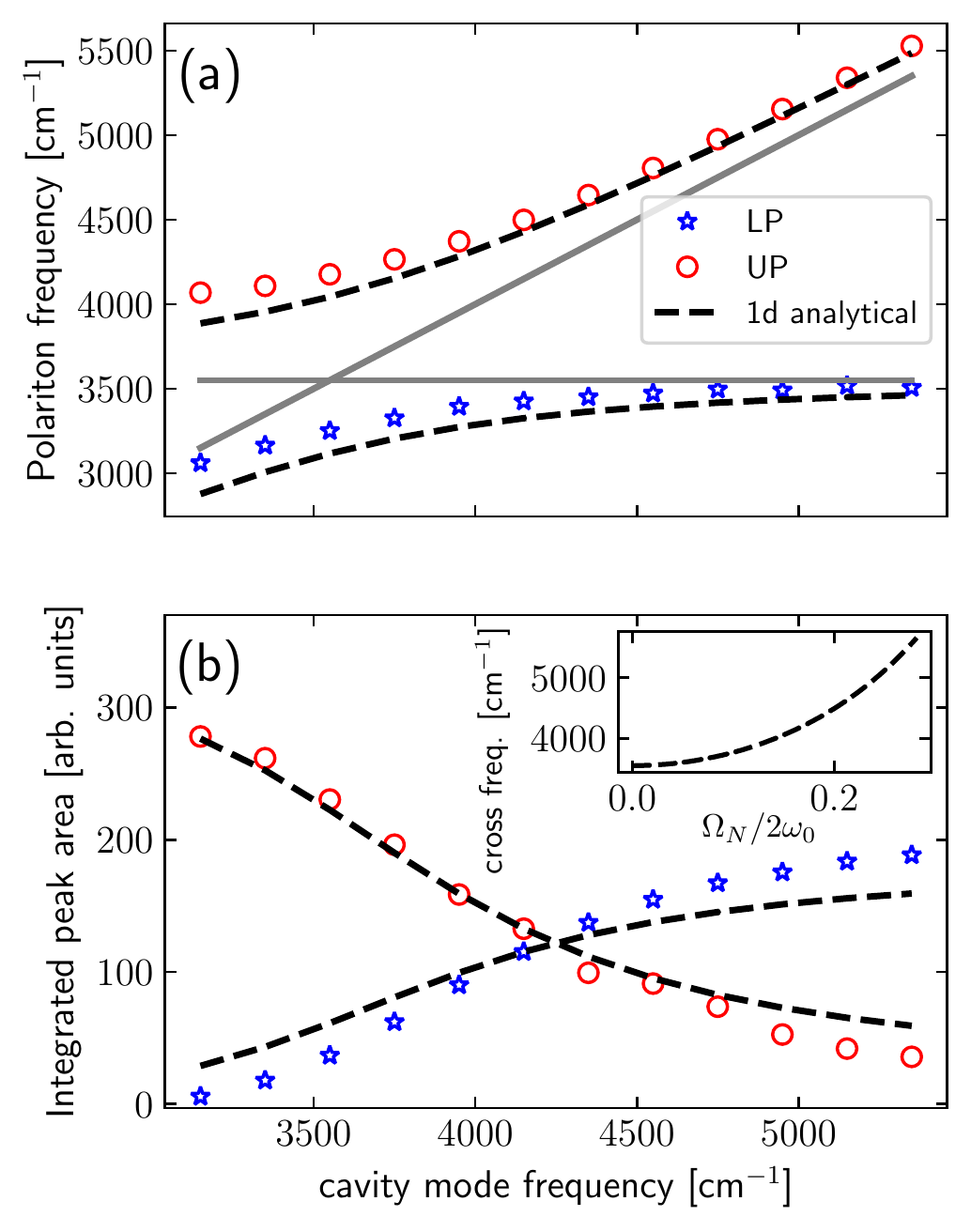}
		\caption{(a) Polariton frequency and (b) integrated peak area of polaritons as a function of the cavity mode frequency. Note that the energy splitting between polaritons is minimized when the cavity mode has frequency $3550 \text{\ cm}^{-1}$, but the upper and lower polaritons become symmetric in intensity at a different cavity mode frequency ($4250 \text{\ cm}^{-1}$). The simulation data  for liquid water (scattered points) are compared with the analytical expressions of the simplified 1D model (black lines, see Eqs. \eqref{eq:omega_LP_UP} and \eqref{eq:I_area_LP_UP}). For simulation parameters, $\widetilde{\varepsilon} = 5\times 10^{-4}$ a.u. and all other parameters are the same as in Fig. \ref{fig:Rabi_splitting}. For parameters of the analytical expressions, we take $\omega_0=3550 \text{\ cm}^{-1}$ and $\Omega_N = 937 \text{\ cm}^{-1}$, which corresponds to the resonant Rabi frequency when $\widetilde{\varepsilon} = 5\times 10^{-4}$ a.u. (see Fig. \ref{fig:Rabi_splitting}a). The gray solid lines in Fig. \ref{fig:Rabi_splitting_freq}a represents the uncoupled \ch{O-H} stretch mode frequency and the cavity mode frequency. The insert in Fig. \ref{fig:Rabi_splitting_freq}b plots the cavity mode frequency corresponding to the case of symmetric polaritons (i.e., the crossing point frequency in Fig. \ref{fig:Rabi_splitting_freq}b) as a function of $\Omega_N/2\omega_0$.}
		\label{fig:Rabi_splitting_freq}
	\end{figure}
	
	For liquid water in the cavity, in Fig. \ref{fig:Rabi_splitting_freq}, we further investigate how (a) the polariton frequencies and (b) the integrated peak areas of polaritons depend on the cavity mode frequency for $\widetilde{\varepsilon} = 5\times 10^{-4}$ a.u., which is well in the USC regime. The simulation data (scatter points) agree well with the analytical result (dashed black lines) for the simplified 1D model (Eqs. \eqref{eq:omega_LP_UP} and \eqref{eq:I_area_LP_UP}). 
	As shown in Fig. \ref{fig:Rabi_splitting_freq}a, the energy difference between the polaritons is minimal at resonance ($\sim 3550 \text{\ cm}^{-1}$), in which the uncoupled \ch{O-H} stretch mode frequency crosses with the cavity mode frequency; see gray solid lines. Such a cross corresponds to the maximally hybridized light-matter state. By contrast, when the cavity mode frequency is larger (smaller) than the molecular frequency, the LP (UP) becomes increasingly dominated by the \ch{O-H} stretch mode (as evident from the uncoupled case for which this mode is represented by the gray horizontal line).
		
	Our model implies that for the uncoupled molecule-cavity case, only the molecular optical transition is coupled to the far field. This suggests that in contrast to the resonance case when the UP peak is larger than the LP peak, when the cavity mode frequency becomes sufficiently large (i.e., the LP is mostly constituted by the matter side), the LP should have a larger peak size than the UP. This finding is confirmed by Fig. \ref{fig:Rabi_splitting_freq}b. More interestingly, Fig. \ref{fig:Rabi_splitting_freq}b also shows the  
	symmetric peak size of polaritons occurs when the cavity mode frequency is $\sim 4250 \text{\ cm}^{-1}$, which is far beyond the \ch{O-H} stretch frequency of liquid water ($\sim 3550 \text{\ cm}^{-1}$). 
	Therefore, in principle, from this fact, one would predict that one can engineer the relative strength of polaritons by tuning the cavity mode frequency.
	Furthermore, the inset of Fig. \ref{fig:Rabi_splitting_freq}b plots the cavity mode frequency (for which the polariton intensities become symmetric) as a function of $\Omega_N/2\omega_0$. Again, we find that for large $\Omega_N/2\omega_0$, detecting polaritons with symmetric intensities requires a very large off-resonant cavity mode frequency.

	\subsection{Static Equilibrium Properties of a Single Molecule}

	\begin{figure}
		\centering
	\includegraphics[width=0.7\linewidth]{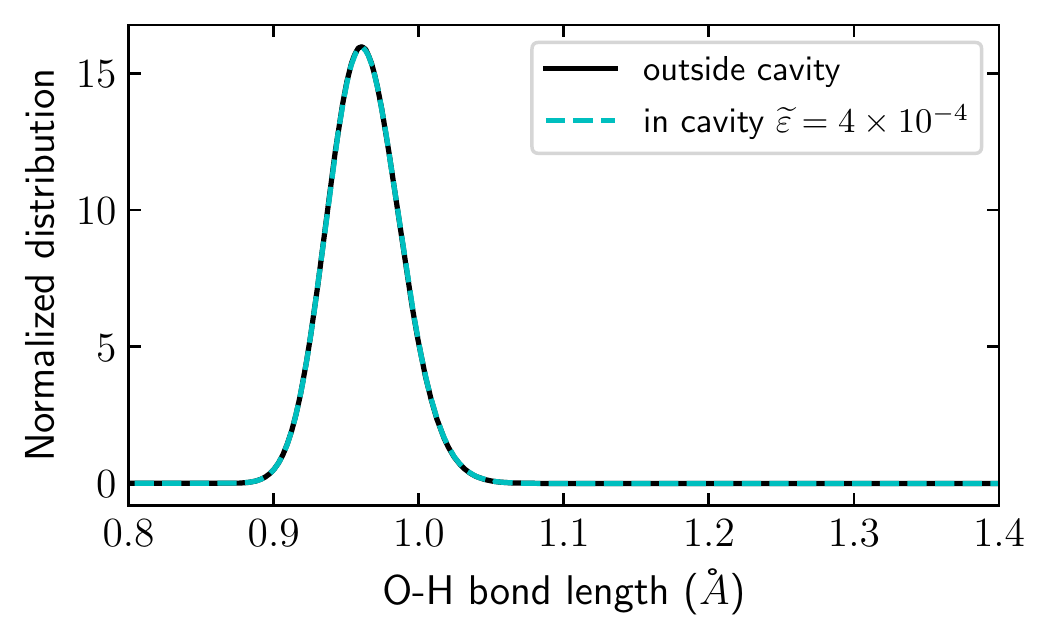}
	\caption{Normalized bond length distribution of \ch{O-H} in liquid water. The result outside the cavity (solid black) is compared with that inside the cavity (with effective coupling strength $\widetilde{\varepsilon} = 4\times 10^{-4}$ a.u., cyan dashed). All other parameters are set the same as Fig. \ref{fig:IR}. Note that the bond length distribution is not changed by VSC or V-USC.}
	\label{fig:bond_dist}
	\end{figure}
	
	\begin{figure}
		\centering
		\includegraphics[width=0.7\linewidth]{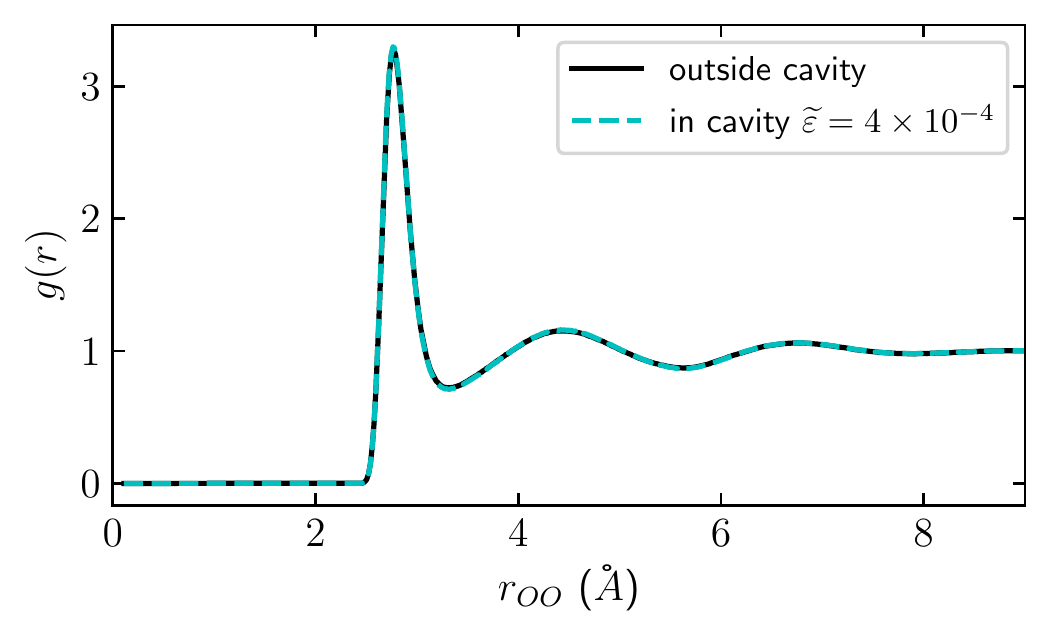}
		\caption{Radical pair distribution function ($g(r)$) of oxygen atoms in liquid water. The result outside the cavity (solid black) is compared with that inside the cavity (with effective coupling strength $\widetilde{\varepsilon} = 4\times 10^{-4}$ a.u., cyan dashed). All other parameters are set the same as Fig. \ref{fig:IR}. Note that $g(r)$ is not changed by VSC or V-USC.}
		\label{fig:pair_dist}
	\end{figure}

	Rabi splitting represents the collective optical response of  liquid water. As shown above, although MD simulations can obtain the IR spectrum of the polaritons in a straightforward way, one can argue that since most important features of the IR spectrum can be qualitatively described by the 1D harmonic model (see SI Appendix, Sec. 2), there is little advantage to perform expensive MD simulations. As has been  argued above, the real advantage of the MD simulations is that one can simultaneously obtain many other physical properties of molecules alongside with the IR spectrum. Below we will investigate whether any property of individual \ch{H2O}  molecules can be changed under VSC or U-VSC.
	
	First, let us consider the static equilibrium properties of \ch{H2O} molecules.
	We recently argued that the classical potential of mean force for a single molecule is not changed by the cavity  \cite{Li2020Origin} under typical VSC or V-USC setups. In fact, with the same proof procedure, it is easy to show that \textit{any} static thermodynamic quantity of the molecules are not changed by the cavity when nuclei and photons are treated classically. This can be illustrated as follows. Given an observable $\mathcal{O} = \mathcal{O}(\{\mathbf{P}_{nj}\}, \{\mathbf{R}_{nj}\})$ which is a function of the molecules only, the thermodynamic average for this variable inside the cavity ($\avg{\mathcal{O}}_{\text{QED}}$) is calculated by
	\begin{subequations}
		\begin{align}
		\avg{\mathcal{O}}_{\text{QED}} &= \frac{\int d\{\mathbf{R}_{nj}\}  d\{\mathbf{P}_{nj}\}
		d \{\widetilde{q}_{k,\lambda}\} d\{\widetilde{p}_{k,\lambda}\} \mathcal{O} e^{-\beta H_{\text{QED}}^{\text{G}}}	
	}{\int d\{\mathbf{R}_{nj}\}  d\{\mathbf{P}_{nj}\}
	d \{\widetilde{q}_{k,\lambda}\} d\{\widetilde{p}_{k,\lambda}\}  e^{-\beta H_{\text{QED}}^{\text{G}}}} \\
	&= \frac{\int d\{\mathbf{R}_{nj}\}  d\{\mathbf{P}_{nj}\}
	\mathcal{O} e^{-\beta H_{\text{M}}^{\text{G}}}	
	}{\int d\{\mathbf{R}_{nj}\}  d\{\mathbf{P}_{nj}\}  e^{-\beta H_{\text{M}}^{\text{G}}}} 
	= \avg{\mathcal{O}}_{\text{M}}
		\end{align}
	\end{subequations}
	which is identical to the average outside the cavity ($\avg{\mathcal{O}}_{\text{M}}$) after the integration over the photon modes, 
	where $H_{\text{QED}}^{\text{G}}$ and $H_{\text{M}}^{\text{G}}$ are defined in SI Appendix, Sec. 1.
	
	Even though the mathematical proof guarantees that the static thermodynamic properties are not changed inside the cavity, it is still very helpful to check some static properties in simulation, as it provides a tool for checking the numerical convergence. Fig. \ref{fig:bond_dist} plots the normalized bond length distribution of the \ch{O-H} bond. Fig. \ref{fig:pair_dist} plots the radical pair distribution function between the oxygen atoms. For these two static properties, the results outside the cavity (solid black)  agree exactly with the results inside the cavity (with effective coupling strength $\widetilde{\varepsilon} = 4\times 10^{-4}$ a.u.). We have checked the results under other coupling strengths and this conclusion is not changed.
	Hence, both analytical and numerical treatments suggest that the static thermodynamic properties are not changed inside the cavity within a classical treatment of nuclei and photons. Of course, quantum effects of nuclei and photons may play a role in the cavity modification of static properties, which needs further investigation.

	\subsection{Dynamical Properties of a Single Molecule}
	
	\begin{figure}
		\centering
	\includegraphics[width=0.8\linewidth]{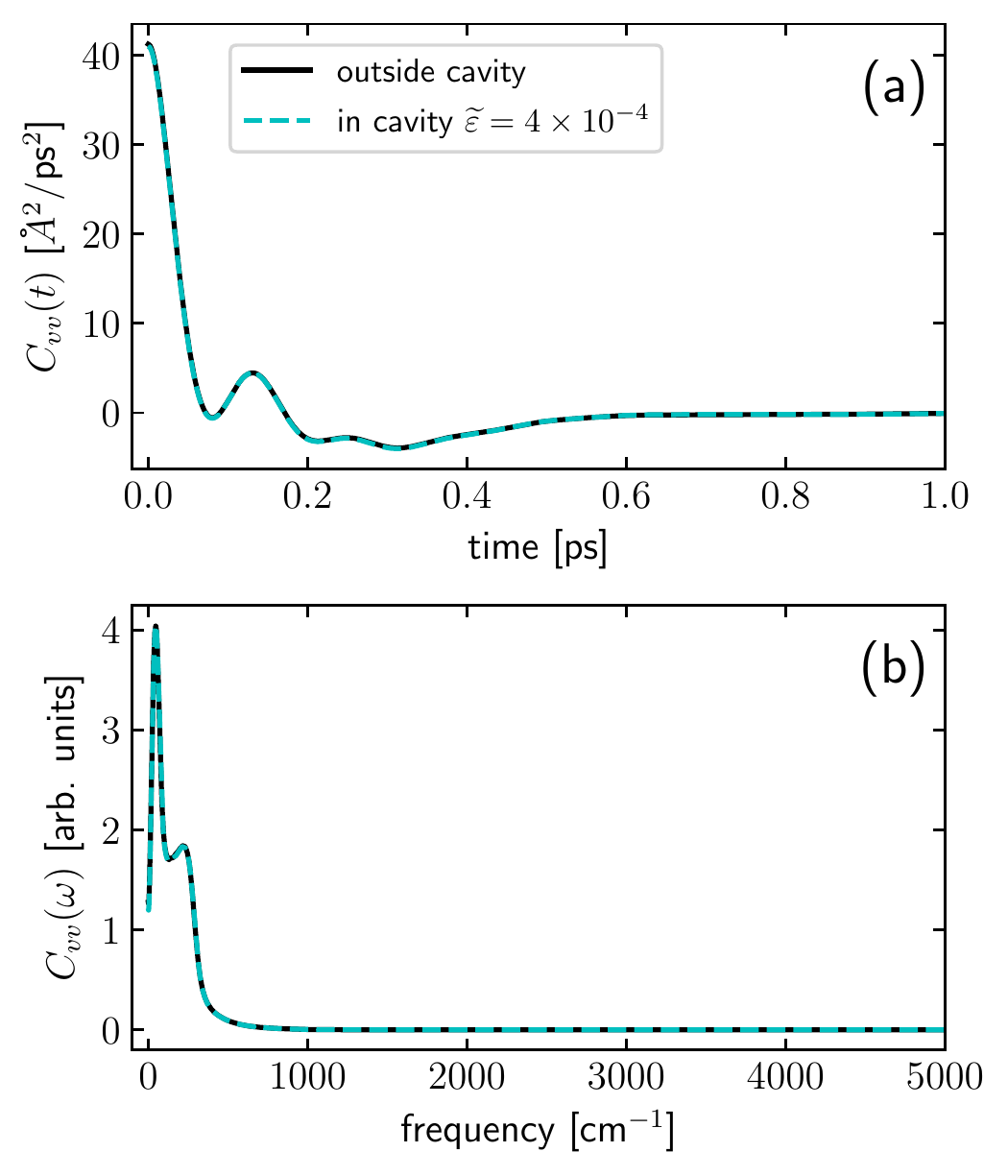}
	\caption{Velocity autocorrelation function (VACF) of the center of mass of individual \ch{H2O} molecules: (a)  the time-domain results and (b)  the corresponding Fourier transform.
	The results outside the cavity (black solid) is compared with those inside the cavity (with effective coupling strength $\widetilde{\varepsilon} = 4\times 10^{-4}$ a.u., cyan dashed). All other parameters are set the same as Fig. \ref{fig:IR}. Note that VACF is not changed by VSC or V-USC.}
	\label{fig:diffusion}
	\end{figure}

	Second, let us move to the dynamical properties of individual \ch{H2O} molecules. In particular, we are interested in whether the translational or rotational motion of a single \ch{H2O} molecule is changed under VSC. 
	
	According to linear response theory, the translational diffusion of \ch{H2O} can be described by the VACF ($C_{vv}(t)$) of the center of mass of each molecule:
	\begin{equation}\label{eq:vacf}
	C_{vv}(t) = \avg{\mathbf{v}(t)\mathbf{v}(0)}
	\end{equation}
	One can calculate the diffusion constant $D$ from $C_{vv}(t)$ by $D = \frac{1}{3} \int_{0}^{+\infty} C_{vv}(t)dt$.
	
\begin{figure}
		\centering
		\includegraphics[width=0.8\linewidth]{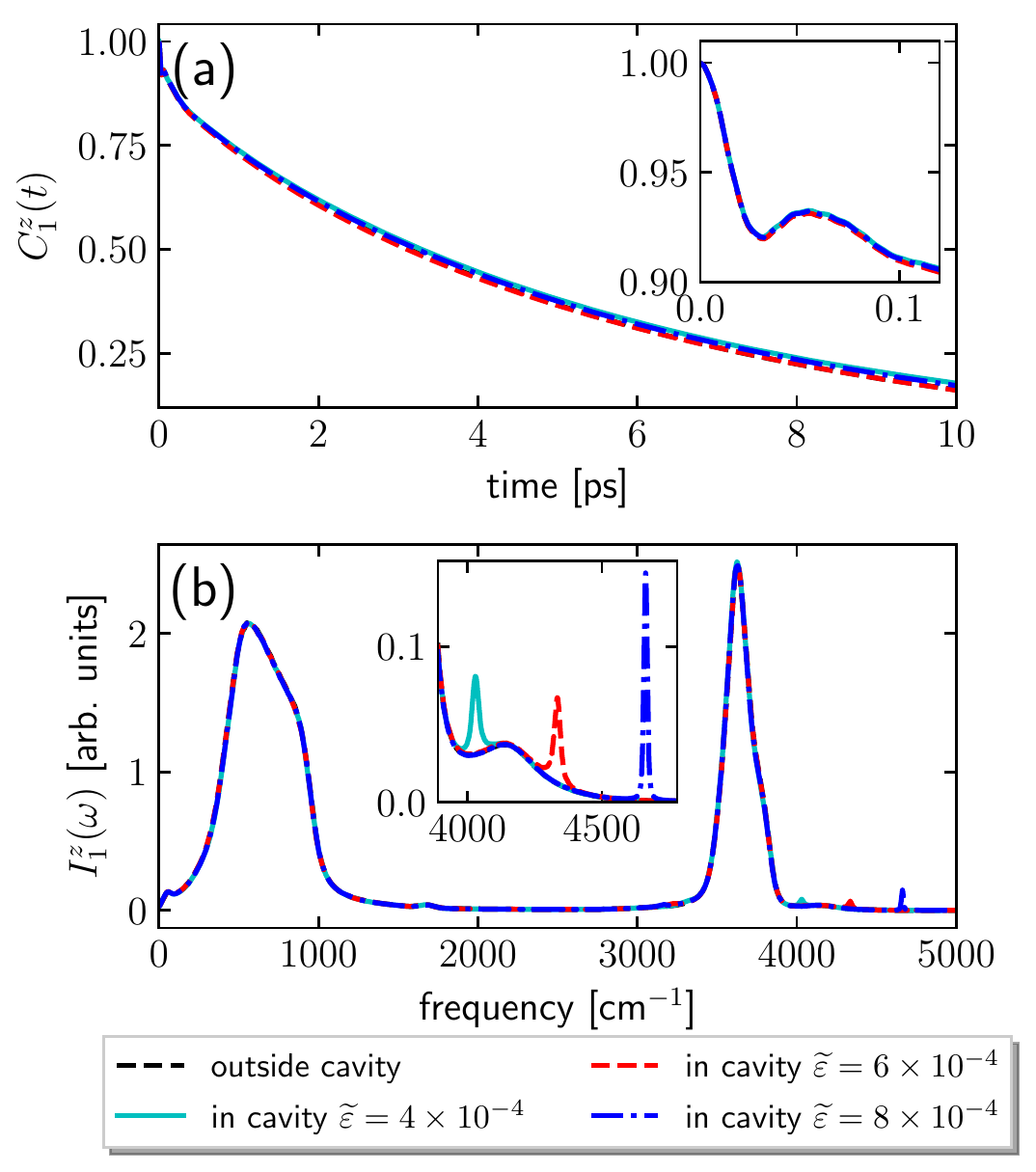}
		\caption{$z$-component of first-order orientational autocorrelation function (OACF) of individual \ch{H2O} molecules. (a) plots the time-domain results ($C_1^z(t)$) and (b) plots the corresponding  spectrum ($I_1^z(\omega) = \omega^2C_1^z(\omega)$).
		A zoom-in inset is also plotted in each subplot.
		The results outside the cavity (black dashed) is compared with those inside the cavity (with effective coupling strength $\widetilde{\varepsilon}$ as $4\times 10^{-4}$ (cyan solid), $6\times 10^{-4}$ (red dashed), and $8\times 10^{-4}$ a.u. (blue dash-dotted), respectively.  All other parameters are set the same as Fig. \ref{fig:IR}. Note that OACF is changed by V-USC.}
		\label{fig:orientation}
	\end{figure}

	Fig. \ref{fig:diffusion}(a) plots $C_{vv}(t)$ as a function of time for the center of mass of \ch{H2O}. The exact agreement between the result outside the cavity (black solid) and that inside the cavity (cyan dotted, with effective coupling strength $\widetilde{\varepsilon} = 4\times 10^{-4}$ a.u.) suggests that $C_{vv}(t)$ is not changed by VSC or V-USC.
	 This finding can also be convinced by looking at the Fourier transform $C_{vv}(\omega)$, which is shown in Fig. \ref{fig:diffusion}b. 
	 Again, we have confirmed this conclusion by checking other coupling strengths. Note that although the VACF for the center of mass motion of \ch{H2O} is not changed by VSC or V-USC, we do find a small cavity modification of the VACF spectrum for the internal modes of individual \ch{H2O} molecules (e.g., the VACF for the \ch{O-H} bond). Such a modification is similar to Fig. \ref{fig:orientation}b but is less intense; see SI Appendix, Sec. 3.
	
	As for the rotational behavior, according to linear response theory, one must compute the orientational autocorrelation function (OACF, denoted by $C_l(t)$)  \cite{Lynden-Bell1981,Impey1982,Miller2005}, which is defined as
	\begin{equation}\label{eq:oacf}
	C_l(t) = \avg{P_l\left[\vu_n(0)\cdot \vu_n(t) \right]}
	\end{equation}
	where $\vu_n(t)$ denotes the three principal inertial axes of molecule $n$ at time $t$, and $P_l$ denotes the Legendre polynomial of index $l$. For simplicity, we will study only the first order of OACF, which means $P_1\left[\vu_n(0)\cdot \vu_n(t)\right] = \vu_n(0)\cdot \vu_n(t)$.
	
	For \ch{H2O}, the $z$ axis of the principal axes coincides with the dipole moment direction. In Fig. \ref{fig:orientation}a, we plot $C_1^{z}(t)$, the $z$-component of the first-order OACF, as a function of time. The inset zooms in the initial rotation relaxation process when time $t < 0.1$ ps. The outside-cavity result (black dashed) largely agrees with results inside cavity [with the effective coupling strength $\widetilde{\varepsilon}$ as $4\times 10^{-4}$ (cyan solid), $6\times 10^{-4}$ (red dashed), $8\times 10^{-4}$  a.u. (blue dash-dotted), respectively]. 
	Fig. \ref{fig:orientation}b plots the corresponding spectrum $I_1^z(\omega)$, which is defined as
	\begin{equation}
	I_1^z(\omega) = \omega^2 C_1^z(\omega)
	\end{equation}
	$I_1^z(\omega)$ can be regarded as the single-molecule IR spectroscopy along the dipole-motion direction, which describes how a single molecule rotates in the environment.
	As clearly shown in the zoom-in inset, for large enough $\widetilde{\varepsilon}$ (in the V-USC limit, or $\widetilde{\varepsilon} \geq 4\times 10^{-4}$ a.u.), an additional small peak  emerges with intensities $2\%\sim 8\%$ of the peak from a  bare molecule.
	Compared with the IR spectrum of the liquid water in Fig. \ref{fig:IR}, these additional small peaks have the same frequencies as the UP peaks, demonstrating the modification of  single-molecule rotation under V-USC.  Note that for smaller $\widetilde{\varepsilon}$ (i.e., in the VSC limit), the additional peak will be covered by the large bare-molecule peak and is hardly identifiable.
	The change of the rotational behavior of individual molecules may possibly change the ground-state chemistry for many scenarios, which should be extensively studied in the future.
	Lastly, we emphasize that apart from these additional peaks, the width of the bare-molecule peaks is mostly unchanged.

	\subsection{Effects of a Multimode Cavity}	
	Note that all the results presented above consider only a single cavity mode frequency, which is valid when the fundamental cavity mode is near resonance with the highest molecular vibrational frequency (i.e., the \ch{O-H} stretch mode $\sim 3500 \text{\ cm}^{-1}$ for liquid water). However, for a cavity with a larger length, the fundamental cavity mode frequency can be much smaller than that of the \ch{O-H} stretch mode. In such a case, many cavity modes must be taken into account.  In SI Appendix Sec. 4, we show the results when liquid water is coupled to a multimode cavity. When different cavity modes are resonantly coupled to the vibrational modes respectively, we observe a multimode Rabi splitting in the IR spectrum, i.e., several Rabi splittings are formed for different vibrational modes. At the same time, however, the above findings regarding the single-molecule properties are not changed when a multimode cavity is considered.

	\section{Conclusion}\label{sec:discussion}
	In conclusion, we have performed classical cavity MD simulations under VSC or V-USC. With liquid water as an example, when the cavity modes are resonantly coupled to the \ch{O-H} stretch mode, we have found asymmetric Rabi splitting of the \ch{O-H} stretch peak in the IR spectrum where the LP is suppressed and the UP is enhanced. Such asymmetry can be inverted  (i.e., the LP is enhanced and the UP is suppressed) by increasing the cavity mode frequency.
	Moreover, with a classical treatment of nuclei and photons, while we have found \textit{no} modification of the static equilibrium properties as well as the translational diffusion of liquid water, we have observed that
	the OACF of \ch{H2O} molecules are modified under V-USC. Such observation may perhaps help  understand the catalytic effect of VSC or V-USC.

	Based on the current framework of cavity MD, future directions should focus on  (i)  path-integral calculations to study  quantum effects in the modification of the molecular dynamical properties; and (ii) \textit{ab initio} cavity MD  simulations of chemical reactions under VSC or V-USC. This cavity MD framework can also be used to simulate recently reported 2D-IR spectroscopy studies \cite{Xiang2018,Xiang2019} on polariton relaxation dynamics.
	At the same time, obtaining analytical solutions of cavity modification of the dynamical properties would also be very helpful.
	We hope such studies will help solve  the mystery of the catalytic effects underlying VSC or V-USC  in the near future.
	 
	\section*{Materials and Methods}\label{sec:methods}
	We calculate several equilibrium and linear response observables of water ($\mathbf{F}_{nj}^{(0)}$, $d_{ng,\lambda}$, and $\partial d_{ng, \lambda}/\partial \mathbf{R}_{nj}$ in  Eq. \eqref{eq:EOM_MD_PBC}, $\vmu_S$ in Eq. \eqref{eq:IR_equation_cavity}, bond length, $g(r)$, VACF in Eq. \eqref{eq:vacf}, and OACF in Eq. \eqref{eq:oacf}) by a classical force field --- the q-TIP4P/F water model  \cite{Habershon2009} --- which provides the simplest description of both the equilibrium and dynamic properties of liquid water. Coupling to an optical cavity mode is included by modifying an open-source MD package I-PI  \cite{Kapil2019}.
		
	As detailed in the SI Appendix Sec. 1, the cavity is placed along the $z$-axis. A pair of thick \ch{SiO2}  layers are placed between the cavity mirrors so that the water molecules can move  freely only in a small region (but still on the order of microns) near the cavity center. Such additional \ch{SiO2} layers are used (i) to ensure the intermolcular interactions between \ch{H2O} molecules are the same as those in free space, and (ii) to validate the long-wave approximation that we have taken from the very beginning. We consider only two cavity modes polarized along $x$ and $y$ directions, both of which are resonant with the \ch{O-H} stretch mode. We set the auxiliary mass for the two photons as $m_{k, \lambda} = 1$ a.u. (atomic units).

	Using periodic boundary condition as detailed in the SI Appendix Sec. 1, we simulate 216 \ch{H2O} molecules in a cubic cell with length $35.233$ a.u., so that the water density is $0.997 \text{\ g cm}^{-1}$. At 300 K, we first run the simulation for 150 ps to guarantee thermal equilibrium under a NVT ensemble where a Langevin thermostat is added on the momenta of all particles (nuclei + photons). The resulting equilibrium configurations are used as starting points for  80 consecutive NVE trajectories of length 20 ps. At the beginning of each trajectory the velocities are resampled by a Maxwell-Boltzman distribution under 300 K. The intermolecular Coulombic interactions are calculated by an Ewald summation.  The simulation step is set as $0.5$ fs and we store the snapshots of trajectories every 2 fs. 
		
	See SI Appendix Sec. 1 for details of the q-TIP4P/F force field and the implementation details. The code and simulation data are available on Github \cite{TELi2020Github}.

	\section*{Acknowledgements}

	This material is based upon work supported by the U.S. Department of Energy, Office of Science, Office of Basic Energy Sciences under Award Number DE-SC0019397 (J.E.S.), U.S. National Science Foundation, Grant No. CHE1665291 (A.N.) and the Israel-U.S. Binational Science Foundation, Grant No. 2014113 (A.N.). T.E.L. also acknowledges the Vagelos Institute for Energy Science and Technology at the University of Pennsylvania for a graduate fellowship.  This research also used resources of the National Energy Research Scientific Computing Center (NERSC), a U.S. Department of Energy Office of Science User Facility operated under Contract No. DE-AC02-05CH11231.


\begin{thebibliography}{55}%
		\makeatletter
		\providecommand \@ifxundefined [1]{%
			\@ifx{#1\undefined}
		}%
		\providecommand \@ifnum [1]{%
			\ifnum #1\expandafter \@firstoftwo
			\else \expandafter \@secondoftwo
			\fi
		}%
		\providecommand \@ifx [1]{%
			\ifx #1\expandafter \@firstoftwo
			\else \expandafter \@secondoftwo
			\fi
		}%
		\providecommand \natexlab [1]{#1}%
		\providecommand \enquote  [1]{``#1''}%
		\providecommand \bibnamefont  [1]{#1}%
		\providecommand \bibfnamefont [1]{#1}%
		\providecommand \citenamefont [1]{#1}%
		\providecommand \href@noop [0]{\@secondoftwo}%
		\providecommand \href [0]{\begingroup \@sanitize@url \@href}%
		\providecommand \@href[1]{\@@startlink{#1}\@@href}%
		\providecommand \@@href[1]{\endgroup#1\@@endlink}%
		\providecommand \@sanitize@url [0]{\catcode `\\12\catcode `\$12\catcode
			`\&12\catcode `\#12\catcode `\^12\catcode `\_12\catcode `\%12\relax}%
		\providecommand \@@startlink[1]{}%
		\providecommand \@@endlink[0]{}%
		\providecommand \url  [0]{\begingroup\@sanitize@url \@url }%
		\providecommand \@url [1]{\endgroup\@href {#1}{\urlprefix }}%
		\providecommand \urlprefix  [0]{URL }%
		\providecommand \Eprint [0]{\href }%
		\providecommand \doibase [0]{http://dx.doi.org/}%
		\providecommand \selectlanguage [0]{\@gobble}%
		\providecommand \bibinfo  [0]{\@secondoftwo}%
		\providecommand \bibfield  [0]{\@secondoftwo}%
		\providecommand \translation [1]{[#1]}%
		\providecommand \BibitemOpen [0]{}%
		\providecommand \bibitemStop [0]{}%
		\providecommand \bibitemNoStop [0]{.\EOS\space}%
		\providecommand \EOS [0]{\spacefactor3000\relax}%
		\providecommand \BibitemShut  [1]{\csname bibitem#1\endcsname}%
		\let\auto@bib@innerbib\@empty
		\bibitem [{\citenamefont {Herrera}\ and\ \citenamefont
			{Owrutsky}(2020)}]{Herrera2019}%
		\BibitemOpen
		\bibfield  {author} {\bibinfo {author} {\bibfnamefont {F.}~\bibnamefont
				{Herrera}}\ and\ \bibinfo {author} {\bibfnamefont {J.}~\bibnamefont
				{Owrutsky}},\ }\href {\doibase 10.1063/1.5136320} {\bibfield  {journal}
			{\bibinfo  {journal} {J. Chem. Phys.}\ }\textbf {\bibinfo {volume} {152}},\
			\bibinfo {pages} {100902} (\bibinfo {year} {2020})}\BibitemShut {NoStop}%
		\bibitem [{\citenamefont {{Frisk Kockum}}\ \emph {et~al.}(2019)\citenamefont
			{{Frisk Kockum}}, \citenamefont {Miranowicz}, \citenamefont {{De Liberato}},
			\citenamefont {Savasta},\ and\ \citenamefont {Nori}}]{FriskKockum2019}%
		\BibitemOpen
		\bibfield  {author} {\bibinfo {author} {\bibfnamefont {A.}~\bibnamefont
				{{Frisk Kockum}}}, \bibinfo {author} {\bibfnamefont {A.}~\bibnamefont
				{Miranowicz}}, \bibinfo {author} {\bibfnamefont {S.}~\bibnamefont {{De
						Liberato}}}, \bibinfo {author} {\bibfnamefont {S.}~\bibnamefont {Savasta}}, \
			and\ \bibinfo {author} {\bibfnamefont {F.}~\bibnamefont {Nori}},\ }\href
		{\doibase 10.1038/s42254-018-0006-2} {\bibfield  {journal} {\bibinfo
				{journal} {Nat. Rev. Phys.}\ }\textbf {\bibinfo {volume} {1}},\ \bibinfo
			{pages} {19} (\bibinfo {year} {2019})}\BibitemShut {NoStop}%
		\bibitem [{\citenamefont {Shalabney}\ \emph {et~al.}(2015)\citenamefont
			{Shalabney}, \citenamefont {George}, \citenamefont {Hutchison}, \citenamefont
			{Pupillo}, \citenamefont {Genet},\ and\ \citenamefont
			{Ebbesen}}]{Shalabney2015}%
		\BibitemOpen
		\bibfield  {author} {\bibinfo {author} {\bibfnamefont {A.}~\bibnamefont
				{Shalabney}}, \bibinfo {author} {\bibfnamefont {J.}~\bibnamefont {George}},
			\bibinfo {author} {\bibfnamefont {J.}~\bibnamefont {Hutchison}}, \bibinfo
			{author} {\bibfnamefont {G.}~\bibnamefont {Pupillo}}, \bibinfo {author}
			{\bibfnamefont {C.}~\bibnamefont {Genet}}, \ and\ \bibinfo {author}
			{\bibfnamefont {T.~W.}\ \bibnamefont {Ebbesen}},\ }\href {\doibase
			10.1038/ncomms6981} {\bibfield  {journal} {\bibinfo  {journal} {Nat.
					Commun.}\ }\textbf {\bibinfo {volume} {6}},\ \bibinfo {pages} {5981}
			(\bibinfo {year} {2015})}\BibitemShut {NoStop}%
		\bibitem [{\citenamefont {George}\ \emph {et~al.}(2015)\citenamefont {George},
			\citenamefont {Shalabney}, \citenamefont {Hutchison}, \citenamefont {Genet},\
			and\ \citenamefont {Ebbesen}}]{George2015}%
		\BibitemOpen
		\bibfield  {author} {\bibinfo {author} {\bibfnamefont {J.}~\bibnamefont
				{George}}, \bibinfo {author} {\bibfnamefont {A.}~\bibnamefont {Shalabney}},
			\bibinfo {author} {\bibfnamefont {J.~A.}\ \bibnamefont {Hutchison}}, \bibinfo
			{author} {\bibfnamefont {C.}~\bibnamefont {Genet}}, \ and\ \bibinfo {author}
			{\bibfnamefont {T.~W.}\ \bibnamefont {Ebbesen}},\ }\href {\doibase
			10.1021/acs.jpclett.5b00204} {\bibfield  {journal} {\bibinfo  {journal} {J.
					Phys. Chem. Lett.}\ }\textbf {\bibinfo {volume} {6}},\ \bibinfo {pages}
			{1027} (\bibinfo {year} {2015})}\BibitemShut {NoStop}%
		\bibitem [{\citenamefont {George}\ \emph {et~al.}(2016)\citenamefont {George},
			\citenamefont {Chervy}, \citenamefont {Shalabney}, \citenamefont {Devaux},
			\citenamefont {Hiura}, \citenamefont {Genet},\ and\ \citenamefont
			{Ebbesen}}]{George2016}%
		\BibitemOpen
		\bibfield  {author} {\bibinfo {author} {\bibfnamefont {J.}~\bibnamefont
				{George}}, \bibinfo {author} {\bibfnamefont {T.}~\bibnamefont {Chervy}},
			\bibinfo {author} {\bibfnamefont {A.}~\bibnamefont {Shalabney}}, \bibinfo
			{author} {\bibfnamefont {E.}~\bibnamefont {Devaux}}, \bibinfo {author}
			{\bibfnamefont {H.}~\bibnamefont {Hiura}}, \bibinfo {author} {\bibfnamefont
				{C.}~\bibnamefont {Genet}}, \ and\ \bibinfo {author} {\bibfnamefont {T.~W.}\
				\bibnamefont {Ebbesen}},\ }\href {\doibase 10.1103/PhysRevLett.117.153601}
		{\bibfield  {journal} {\bibinfo  {journal} {Phys. Rev. Lett.}\ }\textbf
			{\bibinfo {volume} {117}},\ \bibinfo {pages} {153601} (\bibinfo {year}
			{2016})}\BibitemShut {NoStop}%
		\bibitem [{\citenamefont {Thomas}\ \emph {et~al.}(2016)\citenamefont {Thomas},
			\citenamefont {George}, \citenamefont {Shalabney}, \citenamefont {Dryzhakov},
			\citenamefont {Varma}, \citenamefont {Moran}, \citenamefont {Chervy},
			\citenamefont {Zhong}, \citenamefont {Devaux}, \citenamefont {Genet},
			\citenamefont {Hutchison},\ and\ \citenamefont {Ebbesen}}]{Thomas2016}%
		\BibitemOpen
		\bibfield  {author} {\bibinfo {author} {\bibfnamefont {A.}~\bibnamefont
				{Thomas}}, \bibinfo {author} {\bibfnamefont {J.}~\bibnamefont {George}},
			\bibinfo {author} {\bibfnamefont {A.}~\bibnamefont {Shalabney}}, \bibinfo
			{author} {\bibfnamefont {M.}~\bibnamefont {Dryzhakov}}, \bibinfo {author}
			{\bibfnamefont {S.~J.}\ \bibnamefont {Varma}}, \bibinfo {author}
			{\bibfnamefont {J.}~\bibnamefont {Moran}}, \bibinfo {author} {\bibfnamefont
				{T.}~\bibnamefont {Chervy}}, \bibinfo {author} {\bibfnamefont
				{X.}~\bibnamefont {Zhong}}, \bibinfo {author} {\bibfnamefont
				{E.}~\bibnamefont {Devaux}}, \bibinfo {author} {\bibfnamefont
				{C.}~\bibnamefont {Genet}}, \bibinfo {author} {\bibfnamefont {J.~A.}\
				\bibnamefont {Hutchison}}, \ and\ \bibinfo {author} {\bibfnamefont {T.~W.}\
				\bibnamefont {Ebbesen}},\ }\href {\doibase 10.1002/anie.201605504} {\bibfield
			{journal} {\bibinfo  {journal} {Angew. Chemie Int. Ed.}\ }\textbf {\bibinfo
				{volume} {55}},\ \bibinfo {pages} {11462} (\bibinfo {year}
			{2016})}\BibitemShut {NoStop}%
		\bibitem [{\citenamefont {Lather}\ \emph {et~al.}(2019)\citenamefont {Lather},
			\citenamefont {Bhatt}, \citenamefont {Thomas}, \citenamefont {Ebbesen},\ and\
			\citenamefont {George}}]{Lather2019}%
		\BibitemOpen
		\bibfield  {author} {\bibinfo {author} {\bibfnamefont {J.}~\bibnamefont
				{Lather}}, \bibinfo {author} {\bibfnamefont {P.}~\bibnamefont {Bhatt}},
			\bibinfo {author} {\bibfnamefont {A.}~\bibnamefont {Thomas}}, \bibinfo
			{author} {\bibfnamefont {T.~W.}\ \bibnamefont {Ebbesen}}, \ and\ \bibinfo
			{author} {\bibfnamefont {J.}~\bibnamefont {George}},\ }\href {\doibase
			10.1002/anie.201905407} {\bibfield  {journal} {\bibinfo  {journal} {Angew.
					Chemie Int. Ed.}\ }\textbf {\bibinfo {volume} {58}},\ \bibinfo {pages}
			{10635} (\bibinfo {year} {2019})}\BibitemShut {NoStop}%
		\bibitem [{\citenamefont {Hiura}\ \emph {et~al.}(2018)\citenamefont {Hiura},
			\citenamefont {Shalabney},\ and\ \citenamefont {George}}]{Hiura2018}%
		\BibitemOpen
		\bibfield  {author} {\bibinfo {author} {\bibfnamefont {H.}~\bibnamefont
				{Hiura}}, \bibinfo {author} {\bibfnamefont {A.}~\bibnamefont {Shalabney}}, \
			and\ \bibinfo {author} {\bibfnamefont {J.}~\bibnamefont {George}},\ }\href
		{\doibase 10.26434/chemrxiv.7234721.v4} {\  (\bibinfo {year} {2018}),\
			10.26434/chemrxiv.7234721.v4}\BibitemShut {NoStop}%
		\bibitem [{\citenamefont {Thomas}\ \emph {et~al.}(2019)\citenamefont {Thomas},
			\citenamefont {Lethuillier-Karl}, \citenamefont {Nagarajan}, \citenamefont
			{Vergauwe}, \citenamefont {George}, \citenamefont {Chervy}, \citenamefont
			{Shalabney}, \citenamefont {Devaux}, \citenamefont {Genet}, \citenamefont
			{Moran},\ and\ \citenamefont {Ebbesen}}]{Thomas2019_science}%
		\BibitemOpen
		\bibfield  {author} {\bibinfo {author} {\bibfnamefont {A.}~\bibnamefont
				{Thomas}}, \bibinfo {author} {\bibfnamefont {L.}~\bibnamefont
				{Lethuillier-Karl}}, \bibinfo {author} {\bibfnamefont {K.}~\bibnamefont
				{Nagarajan}}, \bibinfo {author} {\bibfnamefont {R.~M.~A.}\ \bibnamefont
				{Vergauwe}}, \bibinfo {author} {\bibfnamefont {J.}~\bibnamefont {George}},
			\bibinfo {author} {\bibfnamefont {T.}~\bibnamefont {Chervy}}, \bibinfo
			{author} {\bibfnamefont {A.}~\bibnamefont {Shalabney}}, \bibinfo {author}
			{\bibfnamefont {E.}~\bibnamefont {Devaux}}, \bibinfo {author} {\bibfnamefont
				{C.}~\bibnamefont {Genet}}, \bibinfo {author} {\bibfnamefont
				{J.}~\bibnamefont {Moran}}, \ and\ \bibinfo {author} {\bibfnamefont {T.~W.}\
				\bibnamefont {Ebbesen}},\ }\href {\doibase 10.1126/science.aau7742}
		{\bibfield  {journal} {\bibinfo  {journal} {Science}\ }\textbf {\bibinfo
				{volume} {363}},\ \bibinfo {pages} {615} (\bibinfo {year}
			{2019})}\BibitemShut {NoStop}%
		\bibitem [{\citenamefont {Vergauwe}\ \emph {et~al.}(2019)\citenamefont
			{Vergauwe}, \citenamefont {Thomas}, \citenamefont {Nagarajan}, \citenamefont
			{Shalabney}, \citenamefont {George}, \citenamefont {Chervy}, \citenamefont
			{Seidel}, \citenamefont {Devaux}, \citenamefont {Torbeev},\ and\
			\citenamefont {Ebbesen}}]{Vergauwe2019}%
		\BibitemOpen
		\bibfield  {author} {\bibinfo {author} {\bibfnamefont {R.~M.~A.}\
				\bibnamefont {Vergauwe}}, \bibinfo {author} {\bibfnamefont {A.}~\bibnamefont
				{Thomas}}, \bibinfo {author} {\bibfnamefont {K.}~\bibnamefont {Nagarajan}},
			\bibinfo {author} {\bibfnamefont {A.}~\bibnamefont {Shalabney}}, \bibinfo
			{author} {\bibfnamefont {J.}~\bibnamefont {George}}, \bibinfo {author}
			{\bibfnamefont {T.}~\bibnamefont {Chervy}}, \bibinfo {author} {\bibfnamefont
				{M.}~\bibnamefont {Seidel}}, \bibinfo {author} {\bibfnamefont
				{E.}~\bibnamefont {Devaux}}, \bibinfo {author} {\bibfnamefont
				{V.}~\bibnamefont {Torbeev}}, \ and\ \bibinfo {author} {\bibfnamefont
				{T.~W.}\ \bibnamefont {Ebbesen}},\ }\href {\doibase 10.1002/anie.201908876}
		{\bibfield  {journal} {\bibinfo  {journal} {Angew. Chemie Int. Ed.}\ }\textbf
			{\bibinfo {volume} {58}},\ \bibinfo {pages} {15324} (\bibinfo {year}
			{2019})}\BibitemShut {NoStop}%
		\bibitem [{\citenamefont {Xiang}\ \emph {et~al.}(2018)\citenamefont {Xiang},
			\citenamefont {Ribeiro}, \citenamefont {Dunkelberger}, \citenamefont {Wang},
			\citenamefont {Li}, \citenamefont {Simpkins}, \citenamefont {Owrutsky},
			\citenamefont {Yuen-Zhou},\ and\ \citenamefont {Xiong}}]{Xiang2018}%
		\BibitemOpen
		\bibfield  {author} {\bibinfo {author} {\bibfnamefont {B.}~\bibnamefont
				{Xiang}}, \bibinfo {author} {\bibfnamefont {R.~F.}\ \bibnamefont {Ribeiro}},
			\bibinfo {author} {\bibfnamefont {A.~D.}\ \bibnamefont {Dunkelberger}},
			\bibinfo {author} {\bibfnamefont {J.}~\bibnamefont {Wang}}, \bibinfo {author}
			{\bibfnamefont {Y.}~\bibnamefont {Li}}, \bibinfo {author} {\bibfnamefont
				{B.~S.}\ \bibnamefont {Simpkins}}, \bibinfo {author} {\bibfnamefont {J.~C.}\
				\bibnamefont {Owrutsky}}, \bibinfo {author} {\bibfnamefont {J.}~\bibnamefont
				{Yuen-Zhou}}, \ and\ \bibinfo {author} {\bibfnamefont {W.}~\bibnamefont
				{Xiong}},\ }\href {\doibase 10.1073/pnas.1722063115} {\bibfield  {journal}
			{\bibinfo  {journal} {Proc. Natl. Acad. Sci.}\ }\textbf {\bibinfo {volume}
				{115}},\ \bibinfo {pages} {4845} (\bibinfo {year} {2018})}\BibitemShut
		{NoStop}%
		\bibitem [{\citenamefont {Xiang}\ \emph {et~al.}(2019)\citenamefont {Xiang},
			\citenamefont {Ribeiro}, \citenamefont {Chen}, \citenamefont {Wang},
			\citenamefont {Du}, \citenamefont {Yuen-Zhou},\ and\ \citenamefont
			{Xiong}}]{Xiang2019}%
		\BibitemOpen
		\bibfield  {author} {\bibinfo {author} {\bibfnamefont {B.}~\bibnamefont
				{Xiang}}, \bibinfo {author} {\bibfnamefont {R.~F.}\ \bibnamefont {Ribeiro}},
			\bibinfo {author} {\bibfnamefont {L.}~\bibnamefont {Chen}}, \bibinfo {author}
			{\bibfnamefont {J.}~\bibnamefont {Wang}}, \bibinfo {author} {\bibfnamefont
				{M.}~\bibnamefont {Du}}, \bibinfo {author} {\bibfnamefont {J.}~\bibnamefont
				{Yuen-Zhou}}, \ and\ \bibinfo {author} {\bibfnamefont {W.}~\bibnamefont
				{Xiong}},\ }\href {\doibase 10.1021/acs.jpca.9b04601} {\bibfield  {journal}
			{\bibinfo  {journal} {J. Phys. Chem. A}\ }\textbf {\bibinfo {volume} {123}},\
			\bibinfo {pages} {5918} (\bibinfo {year} {2019})}\BibitemShut {NoStop}%
		\bibitem [{\citenamefont {Hopfield}(1958)}]{Hopfield1958}%
		\BibitemOpen
		\bibfield  {author} {\bibinfo {author} {\bibfnamefont {J.~J.}\ \bibnamefont
				{Hopfield}},\ }\href {\doibase 10.1103/PhysRev.112.1555} {\bibfield
			{journal} {\bibinfo  {journal} {Phys. Rev.}\ }\textbf {\bibinfo {volume}
				{112}},\ \bibinfo {pages} {1555} (\bibinfo {year} {1958})}\BibitemShut
		{NoStop}%
		\bibitem [{\citenamefont {Hern{\'{a}}ndez}\ and\ \citenamefont
			{Herrera}(2019)}]{Hernandez2019}%
		\BibitemOpen
		\bibfield  {author} {\bibinfo {author} {\bibfnamefont {F.~J.}\ \bibnamefont
				{Hern{\'{a}}ndez}}\ and\ \bibinfo {author} {\bibfnamefont {F.}~\bibnamefont
				{Herrera}},\ }\href {\doibase 10.1063/1.5121426} {\bibfield  {journal}
			{\bibinfo  {journal} {J. Chem. Phys.}\ }\textbf {\bibinfo {volume} {151}},\
			\bibinfo {pages} {144116} (\bibinfo {year} {2019})}\BibitemShut {NoStop}%
		\bibitem [{\citenamefont {Du}\ \emph {et~al.}(2018)\citenamefont {Du},
			\citenamefont {Mart{\'{i}}nez-Mart{\'{i}}nez}, \citenamefont {Ribeiro},
			\citenamefont {Hu}, \citenamefont {Menon},\ and\ \citenamefont
			{Yuen-Zhou}}]{Du2018}%
		\BibitemOpen
		\bibfield  {author} {\bibinfo {author} {\bibfnamefont {M.}~\bibnamefont
				{Du}}, \bibinfo {author} {\bibfnamefont {L.~A.}\ \bibnamefont
				{Mart{\'{i}}nez-Mart{\'{i}}nez}}, \bibinfo {author} {\bibfnamefont {R.~F.}\
				\bibnamefont {Ribeiro}}, \bibinfo {author} {\bibfnamefont {Z.}~\bibnamefont
				{Hu}}, \bibinfo {author} {\bibfnamefont {V.~M.}\ \bibnamefont {Menon}}, \
			and\ \bibinfo {author} {\bibfnamefont {J.}~\bibnamefont {Yuen-Zhou}},\ }\href
		{\doibase 10.1039/C8SC00171E} {\bibfield  {journal} {\bibinfo  {journal}
				{Chem. Sci.}\ }\textbf {\bibinfo {volume} {9}},\ \bibinfo {pages} {6659}
			(\bibinfo {year} {2018})}\BibitemShut {NoStop}%
		\bibitem [{\citenamefont {Rudin}\ and\ \citenamefont
			{Reinecke}(1999)}]{Rudin1999}%
		\BibitemOpen
		\bibfield  {author} {\bibinfo {author} {\bibfnamefont {S.}~\bibnamefont
				{Rudin}}\ and\ \bibinfo {author} {\bibfnamefont {T.~L.}\ \bibnamefont
				{Reinecke}},\ }\href {\doibase 10.1103/PhysRevB.59.10227} {\bibfield
			{journal} {\bibinfo  {journal} {Phys. Rev. B}\ }\textbf {\bibinfo {volume}
				{59}},\ \bibinfo {pages} {10227} (\bibinfo {year} {1999})}\BibitemShut
		{NoStop}%
		\bibitem [{\citenamefont {{F. Ribeiro}}\ \emph {et~al.}(2018)\citenamefont {{F.
					Ribeiro}}, \citenamefont {Dunkelberger}, \citenamefont {Xiang}, \citenamefont
			{Xiong}, \citenamefont {Simpkins}, \citenamefont {Owrutsky},\ and\
			\citenamefont {Yuen-Zhou}}]{F.Ribeiro2018}%
		\BibitemOpen
		\bibfield  {author} {\bibinfo {author} {\bibfnamefont {R.}~\bibnamefont {{F.
						Ribeiro}}}, \bibinfo {author} {\bibfnamefont {A.~D.}\ \bibnamefont
				{Dunkelberger}}, \bibinfo {author} {\bibfnamefont {B.}~\bibnamefont {Xiang}},
			\bibinfo {author} {\bibfnamefont {W.}~\bibnamefont {Xiong}}, \bibinfo
			{author} {\bibfnamefont {B.~S.}\ \bibnamefont {Simpkins}}, \bibinfo {author}
			{\bibfnamefont {J.~C.}\ \bibnamefont {Owrutsky}}, \ and\ \bibinfo {author}
			{\bibfnamefont {J.}~\bibnamefont {Yuen-Zhou}},\ }\href {\doibase
			10.1021/acs.jpclett.8b01176} {\bibfield  {journal} {\bibinfo  {journal} {J.
					Phys. Chem. Lett.}\ }\textbf {\bibinfo {volume} {9}},\ \bibinfo {pages}
			{3766} (\bibinfo {year} {2018})}\BibitemShut {NoStop}%
		\bibitem [{\citenamefont {Galego}\ \emph {et~al.}(2019)\citenamefont {Galego},
			\citenamefont {Climent}, \citenamefont {Garcia-Vidal},\ and\ \citenamefont
			{Feist}}]{Galego2019}%
		\BibitemOpen
		\bibfield  {author} {\bibinfo {author} {\bibfnamefont {J.}~\bibnamefont
				{Galego}}, \bibinfo {author} {\bibfnamefont {C.}~\bibnamefont {Climent}},
			\bibinfo {author} {\bibfnamefont {F.~J.}\ \bibnamefont {Garcia-Vidal}}, \
			and\ \bibinfo {author} {\bibfnamefont {J.}~\bibnamefont {Feist}},\ }\href
		{\doibase 10.1103/PhysRevX.9.021057} {\bibfield  {journal} {\bibinfo
				{journal} {Phys. Rev. X}\ }\textbf {\bibinfo {volume} {9}},\ \bibinfo {pages}
			{021057} (\bibinfo {year} {2019})}\BibitemShut {NoStop}%
		\bibitem [{\citenamefont {Campos-Gonzalez-Angulo}\ \emph
			{et~al.}(2019)\citenamefont {Campos-Gonzalez-Angulo}, \citenamefont
			{Ribeiro},\ and\ \citenamefont {Yuen-Zhou}}]{Campos-Gonzalez-Angulo2019}%
		\BibitemOpen
		\bibfield  {author} {\bibinfo {author} {\bibfnamefont {J.~A.}\ \bibnamefont
				{Campos-Gonzalez-Angulo}}, \bibinfo {author} {\bibfnamefont {R.~F.}\
				\bibnamefont {Ribeiro}}, \ and\ \bibinfo {author} {\bibfnamefont
				{J.}~\bibnamefont {Yuen-Zhou}},\ }\href {\doibase 10.1038/s41467-019-12636-1}
		{\bibfield  {journal} {\bibinfo  {journal} {Nat. Commun.}\ }\textbf {\bibinfo
				{volume} {10}},\ \bibinfo {pages} {4685} (\bibinfo {year}
			{2019})}\BibitemShut {NoStop}%
		\bibitem [{\citenamefont {Hiura}\ and\ \citenamefont
			{Shalabney}()}]{Hiura2019}%
		\BibitemOpen
		\bibfield  {author} {\bibinfo {author} {\bibfnamefont {H.}~\bibnamefont
				{Hiura}}\ and\ \bibinfo {author} {\bibfnamefont {A.}~\bibnamefont
				{Shalabney}},\ }\href {\doibase 10.26434/chemrxiv.9275777.v1} {\
			10.26434/chemrxiv.9275777.v1}\BibitemShut {NoStop}%
		\bibitem [{\citenamefont {Li}\ \emph {et~al.}(2020{\natexlab{a}})\citenamefont
			{Li}, \citenamefont {Nitzan},\ and\ \citenamefont {Subotnik}}]{Li2020Origin}%
		\BibitemOpen
		\bibfield  {author} {\bibinfo {author} {\bibfnamefont {T.~E.}\ \bibnamefont
				{Li}}, \bibinfo {author} {\bibfnamefont {A.}~\bibnamefont {Nitzan}}, \ and\
			\bibinfo {author} {\bibfnamefont {J.~E.}\ \bibnamefont {Subotnik}},\ }\href
		{http://arxiv.org/abs/2002.09977} {\  (\bibinfo {year}
			{2020}{\natexlab{a}})},\ \Eprint {http://arxiv.org/abs/2002.09977}
		{arXiv:2002.09977} \BibitemShut {NoStop}%
		\bibitem [{\citenamefont {Campos-Gonzalez-Angulo}\ and\ \citenamefont
			{Yuen-Zhou}(2020)}]{Campos-Gonzalez-Angulo2020}%
		\BibitemOpen
		\bibfield  {author} {\bibinfo {author} {\bibfnamefont {J.~A.}\ \bibnamefont
				{Campos-Gonzalez-Angulo}}\ and\ \bibinfo {author} {\bibfnamefont
				{J.}~\bibnamefont {Yuen-Zhou}},\ }\href {\doibase 10.1063/5.0007547}
		{\bibfield  {journal} {\bibinfo  {journal} {J. Chem. Phys.}\ }\textbf
			{\bibinfo {volume} {152}},\ \bibinfo {pages} {161101} (\bibinfo {year}
			{2020})}\BibitemShut {NoStop}%
		\bibitem [{\citenamefont {Zhdanov}(2020)}]{Zhdanov2020}%
		\BibitemOpen
		\bibfield  {author} {\bibinfo {author} {\bibfnamefont {V.~P.}\ \bibnamefont
				{Zhdanov}},\ }\href {\doibase 10.1016/j.chemphys.2020.110767} {\bibfield
			{journal} {\bibinfo  {journal} {Chem. Phys.}\ }\textbf {\bibinfo {volume}
				{535}},\ \bibinfo {pages} {110767} (\bibinfo {year} {2020})}\BibitemShut
		{NoStop}%
		\bibitem [{\citenamefont {Goto}\ and\ \citenamefont
			{Ichimura}(2005)}]{Goto2005}%
		\BibitemOpen
		\bibfield  {author} {\bibinfo {author} {\bibfnamefont {H.}~\bibnamefont
				{Goto}}\ and\ \bibinfo {author} {\bibfnamefont {K.}~\bibnamefont
				{Ichimura}},\ }\href {\doibase 10.1103/PhysRevA.72.054301} {\bibfield
			{journal} {\bibinfo  {journal} {Phys. Rev. A}\ }\textbf {\bibinfo {volume}
				{72}},\ \bibinfo {pages} {054301} (\bibinfo {year} {2005})}\BibitemShut
		{NoStop}%
		\bibitem [{\citenamefont {Li}\ \emph {et~al.}(2020{\natexlab{b}})\citenamefont
			{Li}, \citenamefont {Chen}, \citenamefont {Nitzan},\ and\ \citenamefont
			{Subotnik}}]{Li2020Quasi}%
		\BibitemOpen
		\bibfield  {author} {\bibinfo {author} {\bibfnamefont {T.~E.}\ \bibnamefont
				{Li}}, \bibinfo {author} {\bibfnamefont {H.-T.}\ \bibnamefont {Chen}},
			\bibinfo {author} {\bibfnamefont {A.}~\bibnamefont {Nitzan}}, \ and\ \bibinfo
			{author} {\bibfnamefont {J.~E.}\ \bibnamefont {Subotnik}},\ }\href {\doibase
			10.1103/PhysRevA.101.033831} {\bibfield  {journal} {\bibinfo  {journal}
				{Phys. Rev. A}\ }\textbf {\bibinfo {volume} {101}},\ \bibinfo {pages}
			{033831} (\bibinfo {year} {2020}{\natexlab{b}})}\BibitemShut {NoStop}%
		\bibitem [{\citenamefont {Hoffmann}\ \emph {et~al.}(2019)\citenamefont
			{Hoffmann}, \citenamefont {Sch{\"{a}}fer}, \citenamefont {S{\"{a}}kkinen},
			\citenamefont {Rubio}, \citenamefont {Appel},\ and\ \citenamefont
			{Kelly}}]{Hoffmann2019Benchmark}%
		\BibitemOpen
		\bibfield  {author} {\bibinfo {author} {\bibfnamefont {N.~M.}\ \bibnamefont
				{Hoffmann}}, \bibinfo {author} {\bibfnamefont {C.}~\bibnamefont
				{Sch{\"{a}}fer}}, \bibinfo {author} {\bibfnamefont {N.}~\bibnamefont
				{S{\"{a}}kkinen}}, \bibinfo {author} {\bibfnamefont {A.}~\bibnamefont
				{Rubio}}, \bibinfo {author} {\bibfnamefont {H.}~\bibnamefont {Appel}}, \ and\
			\bibinfo {author} {\bibfnamefont {A.}~\bibnamefont {Kelly}},\ }\href
		{\doibase 10.1063/1.5128076} {\bibfield  {journal} {\bibinfo  {journal} {J.
					Chem. Phys.}\ }\textbf {\bibinfo {volume} {151}},\ \bibinfo {pages} {244113}
			(\bibinfo {year} {2019})}\BibitemShut {NoStop}%
		\bibitem [{\citenamefont {Santhosh}\ \emph {et~al.}(2016)\citenamefont
			{Santhosh}, \citenamefont {Bitton}, \citenamefont {Chuntonov},\ and\
			\citenamefont {Haran}}]{Santhosh2016}%
		\BibitemOpen
		\bibfield  {author} {\bibinfo {author} {\bibfnamefont {K.}~\bibnamefont
				{Santhosh}}, \bibinfo {author} {\bibfnamefont {O.}~\bibnamefont {Bitton}},
			\bibinfo {author} {\bibfnamefont {L.}~\bibnamefont {Chuntonov}}, \ and\
			\bibinfo {author} {\bibfnamefont {G.}~\bibnamefont {Haran}},\ }\href
		{\doibase 10.1038/ncomms11823} {\bibfield  {journal} {\bibinfo  {journal}
				{Nat. Commun.}\ }\textbf {\bibinfo {volume} {7}},\ \bibinfo {pages}
			{ncomms11823} (\bibinfo {year} {2016})}\BibitemShut {NoStop}%
		\bibitem [{\citenamefont {Sukharev}\ and\ \citenamefont
			{Pachter}(2018)}]{Sukharev2018}%
		\BibitemOpen
		\bibfield  {author} {\bibinfo {author} {\bibfnamefont {M.}~\bibnamefont
				{Sukharev}}\ and\ \bibinfo {author} {\bibfnamefont {R.}~\bibnamefont
				{Pachter}},\ }\href {\doibase 10.1063/1.5019953} {\bibfield  {journal}
			{\bibinfo  {journal} {J. Chem. Phys.}\ }\textbf {\bibinfo {volume} {148}},\
			\bibinfo {pages} {094701} (\bibinfo {year} {2018})}\BibitemShut {NoStop}%
		\bibitem [{\citenamefont {Flick}\ \emph {et~al.}(2017)\citenamefont {Flick},
			\citenamefont {Ruggenthaler}, \citenamefont {Appel},\ and\ \citenamefont
			{Rubio}}]{Flick2017}%
		\BibitemOpen
		\bibfield  {author} {\bibinfo {author} {\bibfnamefont {J.}~\bibnamefont
				{Flick}}, \bibinfo {author} {\bibfnamefont {M.}~\bibnamefont {Ruggenthaler}},
			\bibinfo {author} {\bibfnamefont {H.}~\bibnamefont {Appel}}, \ and\ \bibinfo
			{author} {\bibfnamefont {A.}~\bibnamefont {Rubio}},\ }\href {\doibase
			10.1073/pnas.1615509114} {\bibfield  {journal} {\bibinfo  {journal} {Proc.
					Natl. Acad. Sci.}\ }\textbf {\bibinfo {volume} {114}},\ \bibinfo {pages}
			{3026} (\bibinfo {year} {2017})}\BibitemShut {NoStop}%
		\bibitem [{\citenamefont {Luk}\ \emph {et~al.}(2017)\citenamefont {Luk},
			\citenamefont {Feist}, \citenamefont {Toppari},\ and\ \citenamefont
			{Groenhof}}]{Luk2017}%
		\BibitemOpen
		\bibfield  {author} {\bibinfo {author} {\bibfnamefont {H.~L.}\ \bibnamefont
				{Luk}}, \bibinfo {author} {\bibfnamefont {J.}~\bibnamefont {Feist}}, \bibinfo
			{author} {\bibfnamefont {J.~J.}\ \bibnamefont {Toppari}}, \ and\ \bibinfo
			{author} {\bibfnamefont {G.}~\bibnamefont {Groenhof}},\ }\href {\doibase
			10.1021/acs.jctc.7b00388} {\bibfield  {journal} {\bibinfo  {journal} {J.
					Chem. Theory Comput.}\ }\textbf {\bibinfo {volume} {13}},\ \bibinfo {pages}
			{4324} (\bibinfo {year} {2017})}\BibitemShut {NoStop}%
		\bibitem [{\citenamefont {Groenhof}\ \emph {et~al.}(2019)\citenamefont
			{Groenhof}, \citenamefont {Climent}, \citenamefont {Feist}, \citenamefont
			{Morozov},\ and\ \citenamefont {Toppari}}]{Groenhof2019}%
		\BibitemOpen
		\bibfield  {author} {\bibinfo {author} {\bibfnamefont {G.}~\bibnamefont
				{Groenhof}}, \bibinfo {author} {\bibfnamefont {C.}~\bibnamefont {Climent}},
			\bibinfo {author} {\bibfnamefont {J.}~\bibnamefont {Feist}}, \bibinfo
			{author} {\bibfnamefont {D.}~\bibnamefont {Morozov}}, \ and\ \bibinfo
			{author} {\bibfnamefont {J.~J.}\ \bibnamefont {Toppari}},\ }\href {\doibase
			10.1021/acs.jpclett.9b02192} {\bibfield  {journal} {\bibinfo  {journal} {J.
					Phys. Chem. Lett.}\ }\textbf {\bibinfo {volume} {10}},\ \bibinfo {pages}
			{5476} (\bibinfo {year} {2019})}\BibitemShut {NoStop}%
		\bibitem [{\citenamefont {Hiura}\ \emph {et~al.}(2019)\citenamefont {Hiura},
			\citenamefont {Shalabney},\ and\ \citenamefont {George}}]{Hiura2019water}%
		\BibitemOpen
		\bibfield  {author} {\bibinfo {author} {\bibfnamefont {H.}~\bibnamefont
				{Hiura}}, \bibinfo {author} {\bibfnamefont {A.}~\bibnamefont {Shalabney}}, \
			and\ \bibinfo {author} {\bibfnamefont {J.}~\bibnamefont {George}},\ }\href
		{\doibase 10.26434/CHEMRXIV.9808508.V1} {\  (\bibinfo {year} {2019}),\
			10.26434/CHEMRXIV.9808508.V1}\BibitemShut {NoStop}%
		\bibitem [{\citenamefont {Abascal}\ and\ \citenamefont
			{Vega}(2005)}]{Abascal2005}%
		\BibitemOpen
		\bibfield  {author} {\bibinfo {author} {\bibfnamefont {J.~L.~F.}\
				\bibnamefont {Abascal}}\ and\ \bibinfo {author} {\bibfnamefont
				{C.}~\bibnamefont {Vega}},\ }\href {\doibase 10.1063/1.2121687} {\bibfield
			{journal} {\bibinfo  {journal} {J. Chem. Phys.}\ }\textbf {\bibinfo {volume}
				{123}},\ \bibinfo {pages} {234505} (\bibinfo {year} {2005})}\BibitemShut
		{NoStop}%
		\bibitem [{\citenamefont {Habershon}\ and\ \citenamefont
			{Manolopoulos}(2009)}]{Habershon2009}%
		\BibitemOpen
		\bibfield  {author} {\bibinfo {author} {\bibfnamefont {S.}~\bibnamefont
				{Habershon}}\ and\ \bibinfo {author} {\bibfnamefont {D.~E.}\ \bibnamefont
				{Manolopoulos}},\ }\href {\doibase 10.1063/1.3276109} {\bibfield  {journal}
			{\bibinfo  {journal} {J. Chem. Phys.}\ }\textbf {\bibinfo {volume} {131}},\
			\bibinfo {pages} {244518} (\bibinfo {year} {2009})}\BibitemShut {NoStop}%
		\bibitem [{\citenamefont {Corcelli}\ \emph {et~al.}(2004)\citenamefont
			{Corcelli}, \citenamefont {Lawrence},\ and\ \citenamefont
			{Skinner}}]{Corcelli2004}%
		\BibitemOpen
		\bibfield  {author} {\bibinfo {author} {\bibfnamefont {S.~A.}\ \bibnamefont
				{Corcelli}}, \bibinfo {author} {\bibfnamefont {C.~P.}\ \bibnamefont
				{Lawrence}}, \ and\ \bibinfo {author} {\bibfnamefont {J.~L.}\ \bibnamefont
				{Skinner}},\ }\href {\doibase 10.1063/1.1683072} {\bibfield  {journal}
			{\bibinfo  {journal} {J. Chem. Phys.}\ }\textbf {\bibinfo {volume} {120}},\
			\bibinfo {pages} {8107} (\bibinfo {year} {2004})}\BibitemShut {NoStop}%
		\bibitem [{\citenamefont {Cohen-Tannoudji}\ \emph {et~al.}(1997)\citenamefont
			{Cohen-Tannoudji}, \citenamefont {Dupont-Roc},\ and\ \citenamefont
			{Grynberg}}]{Cohen-Tannoudji1997}%
		\BibitemOpen
		\bibfield  {author} {\bibinfo {author} {\bibfnamefont {C.}~\bibnamefont
				{Cohen-Tannoudji}}, \bibinfo {author} {\bibfnamefont {J.}~\bibnamefont
				{Dupont-Roc}}, \ and\ \bibinfo {author} {\bibfnamefont {G.}~\bibnamefont
				{Grynberg}},\ }\href
		{http://www.amazon.de/Photons-Atoms-Introduction-Electrodynamics-Professional/dp/0471184330/ref=sr{\_}1{\_}3?ie=UTF8{\&}qid=1455100150{\&}sr=8-3{\&}keywords=Cohen-Tannoudji+photons}
		{\emph {\bibinfo {title} {{Photons and Atoms: Introduction to Quantum
						Electrodynamics}}}}\ (\bibinfo  {publisher} {Wiley},\ \bibinfo {address} {New
			York},\ \bibinfo {year} {1997})\ pp.\ \bibinfo {pages} {280--295}\BibitemShut
		{NoStop}%
		\bibitem [{\citenamefont {Rokaj}\ \emph {et~al.}(2018)\citenamefont {Rokaj},
			\citenamefont {Welakuh}, \citenamefont {Ruggenthaler},\ and\ \citenamefont
			{Rubio}}]{Rokaj2018}%
		\BibitemOpen
		\bibfield  {author} {\bibinfo {author} {\bibfnamefont {V.}~\bibnamefont
				{Rokaj}}, \bibinfo {author} {\bibfnamefont {D.~M.}\ \bibnamefont {Welakuh}},
			\bibinfo {author} {\bibfnamefont {M.}~\bibnamefont {Ruggenthaler}}, \ and\
			\bibinfo {author} {\bibfnamefont {A.}~\bibnamefont {Rubio}},\ }\href
		{\doibase 10.1088/1361-6455/aa9c99} {\bibfield  {journal} {\bibinfo
				{journal} {J. Phys. B At. Mol. Opt. Phys.}\ }\textbf {\bibinfo {volume}
				{51}},\ \bibinfo {pages} {034005} (\bibinfo {year} {2018})}\BibitemShut
		{NoStop}%
		\bibitem [{\citenamefont {Sch{\"{a}}fer}\ \emph {et~al.}(2020)\citenamefont
			{Sch{\"{a}}fer}, \citenamefont {Ruggenthaler}, \citenamefont {Rokaj},\ and\
			\citenamefont {Rubio}}]{Schafer2020}%
		\BibitemOpen
		\bibfield  {author} {\bibinfo {author} {\bibfnamefont {C.}~\bibnamefont
				{Sch{\"{a}}fer}}, \bibinfo {author} {\bibfnamefont {M.}~\bibnamefont
				{Ruggenthaler}}, \bibinfo {author} {\bibfnamefont {V.}~\bibnamefont {Rokaj}},
			\ and\ \bibinfo {author} {\bibfnamefont {A.}~\bibnamefont {Rubio}},\ }\href
		{\doibase 10.1021/acsphotonics.9b01649} {\bibfield  {journal} {\bibinfo
				{journal} {ACS Photonics}\ }\textbf {\bibinfo {volume} {7}},\ \bibinfo
			{pages} {975} (\bibinfo {year} {2020})}\BibitemShut {NoStop}%
		\bibitem [{\citenamefont {Hoffmann}\ \emph {et~al.}(2020)\citenamefont
			{Hoffmann}, \citenamefont {Lacombe}, \citenamefont {Rubio},\ and\
			\citenamefont {Maitra}}]{Hoffmann2020}%
		\BibitemOpen
		\bibfield  {author} {\bibinfo {author} {\bibfnamefont {N.~M.}\ \bibnamefont
				{Hoffmann}}, \bibinfo {author} {\bibfnamefont {L.}~\bibnamefont {Lacombe}},
			\bibinfo {author} {\bibfnamefont {A.}~\bibnamefont {Rubio}}, \ and\ \bibinfo
			{author} {\bibfnamefont {N.~T.}\ \bibnamefont {Maitra}},\ }\href
		{http://arxiv.org/abs/2001.07330} {\  (\bibinfo {year} {2020})},\ \Eprint
		{http://arxiv.org/abs/2001.07330} {arXiv:2001.07330} \BibitemShut {NoStop}%
		\bibitem [{\citenamefont {Takae}\ and\ \citenamefont
			{Onuki}(2013)}]{Takae2013}%
		\BibitemOpen
		\bibfield  {author} {\bibinfo {author} {\bibfnamefont {K.}~\bibnamefont
				{Takae}}\ and\ \bibinfo {author} {\bibfnamefont {A.}~\bibnamefont {Onuki}},\
		}\href {\doibase 10.1063/1.4821085} {\bibfield  {journal} {\bibinfo
				{journal} {J. Chem. Phys.}\ }\textbf {\bibinfo {volume} {139}},\ \bibinfo
			{pages} {124108} (\bibinfo {year} {2013})}\BibitemShut {NoStop}%
		\bibitem [{\citenamefont {{De Bernardis}}\ \emph {et~al.}(2018)\citenamefont
			{{De Bernardis}}, \citenamefont {Jaako},\ and\ \citenamefont
			{Rabl}}]{DeBernardis2018}%
		\BibitemOpen
		\bibfield  {author} {\bibinfo {author} {\bibfnamefont {D.}~\bibnamefont {{De
						Bernardis}}}, \bibinfo {author} {\bibfnamefont {T.}~\bibnamefont {Jaako}}, \
			and\ \bibinfo {author} {\bibfnamefont {P.}~\bibnamefont {Rabl}},\ }\href
		{\doibase 10.1103/PhysRevA.97.043820} {\bibfield  {journal} {\bibinfo
				{journal} {Phys. Rev. A}\ }\textbf {\bibinfo {volume} {97}},\ \bibinfo
			{pages} {043820} (\bibinfo {year} {2018})}\BibitemShut {NoStop}%
		\bibitem [{foo({\natexlab{a}})}]{footnote_joe}%
		\BibitemOpen
		\href@noop {} {}\bibinfo {howpublished} {In principle, when one considers the
			exact quantum Hamiltonian for systems with light-matter interactions, all (i)
			instantaneous interactions between molecules are canceled exactly by the
			presence of terms that involve (ii) the non-local (and also instantaneous)
			self-interaction of delocalized photon modes. This exact cancellation allows
			for causality to be enforced, such that all meaningful intermolecular
			interactions are carried exclusively the transverse photon field at the speed
			of light. In the present paper, we do not worry about causality and so we
			have ignored the details of the cancellation alluded to above, i.e. we do not
			address how this cancellation is affected by the cavity and the presence of a
			finite number of cavity modes. In principle, the presence of a cavity leads
			to a dressed $\hat{V}_{\text{inter}}^{(nl)}$, i.e. a dressed intermolecular
			interaction (with image charges), and such effects are well understood within
			QED \cite{Takae2013,DeBernardis2018}. However, there is a caveat to this last
			point: an exact expression for $\hat{V}_{\text{inter}}^{(nl)}$ would require
			that we treat all for all EM cavity modes correctly, and the resulting
			$\hat{V}_{\text{inter}}^{(nl)}$ will be complex and exceedingly difficult to
			implement computationally. In practice, we assume that one long-wavelength
			cavity mode that is resonant with the \ch{O-H} mode can be treated
			explicitly, while all modes of higher frequencies are taken as part of the
			environment. For such a prescription, there is no simple means to address the
			correct $\hat{V}_{\text{inter}}^{(nl)}$; however, given how long the length
			scales are (microns), the correctly dressed $\hat{V}_{\text{inter}}^{(nl)}$
			cannot be very different from the standard form of
			$\hat{V}_{\text{inter}}^{(nl)}$. For all of these reasons, we have chosen in
			the present manuscript to work with the standard form of the intermolecular
			interactions ($\hat{V}_{\text{inter}}^{(nl)}$), knowing full well that our
			Hamiltonian slightly double counts some light-matter interactions.}
		({\natexlab{a}})\BibitemShut {NoStop}%
		\bibitem [{\citenamefont {McQuarrie}(1976)}]{McQuarrie1976}%
		\BibitemOpen
		\bibfield  {author} {\bibinfo {author} {\bibfnamefont {D.~A.}\ \bibnamefont
				{McQuarrie}},\ }\href@noop {} {\emph {\bibinfo {title} {{Statistical
						Mechanics}}}}\ (\bibinfo  {publisher} {Harper-Collins Publish- ers},\
		\bibinfo {address} {New York},\ \bibinfo {year} {1976})\BibitemShut {NoStop}%
		\bibitem [{\citenamefont {Gaigeot}\ and\ \citenamefont
			{Sprik}(2003)}]{Gaigeot2003}%
		\BibitemOpen
		\bibfield  {author} {\bibinfo {author} {\bibfnamefont {M.-P.}\ \bibnamefont
				{Gaigeot}}\ and\ \bibinfo {author} {\bibfnamefont {M.}~\bibnamefont
				{Sprik}},\ }\href {\doibase 10.1021/jp034788u} {\bibfield  {journal}
			{\bibinfo  {journal} {J. Phys. Chem. B}\ }\textbf {\bibinfo {volume} {107}},\
			\bibinfo {pages} {10344} (\bibinfo {year} {2003})}\BibitemShut {NoStop}%
		\bibitem [{\citenamefont {Habershon}\ \emph {et~al.}(2008)\citenamefont
			{Habershon}, \citenamefont {Fanourgakis},\ and\ \citenamefont
			{Manolopoulos}}]{Habershon2008}%
		\BibitemOpen
		\bibfield  {author} {\bibinfo {author} {\bibfnamefont {S.}~\bibnamefont
				{Habershon}}, \bibinfo {author} {\bibfnamefont {G.~S.}\ \bibnamefont
				{Fanourgakis}}, \ and\ \bibinfo {author} {\bibfnamefont {D.~E.}\ \bibnamefont
				{Manolopoulos}},\ }\href {\doibase 10.1063/1.2968555} {\bibfield  {journal}
			{\bibinfo  {journal} {J. Chem. Phys.}\ }\textbf {\bibinfo {volume} {129}},\
			\bibinfo {pages} {074501} (\bibinfo {year} {2008})}\BibitemShut {NoStop}%
		\bibitem [{\citenamefont {Nitzan}(2006)}]{Nitzan2006}%
		\BibitemOpen
		\bibfield  {author} {\bibinfo {author} {\bibfnamefont {A.}~\bibnamefont
				{Nitzan}},\ }\href@noop {} {\emph {\bibinfo {title} {{Chemical Dynamics in
						Condensed Phases: Relaxation, Transfer and Reactions in Condensed Molecular
						Systems}}}}\ (\bibinfo  {publisher} {Oxford University Press},\ \bibinfo
		{address} {New York},\ \bibinfo {year} {2006})\BibitemShut {NoStop}%
		\bibitem [{foo({\natexlab{b}})}]{footnote_abe}%
		\BibitemOpen
		\href@noop {} {}\bibinfo {howpublished} {This expression reflects one of
			several suggestions that were made for a correction factor which relates the
			quantum time-correlation function to its classical counterparts
			\cite{Gaigeot2003}.} ({\natexlab{b}})\BibitemShut {NoStop}%
		\bibitem [{\citenamefont {Houdr{\'{e}}}\ \emph {et~al.}(1996)\citenamefont
			{Houdr{\'{e}}}, \citenamefont {Stanley},\ and\ \citenamefont
			{Ilegems}}]{Houdre1996}%
		\BibitemOpen
		\bibfield  {author} {\bibinfo {author} {\bibfnamefont {R.}~\bibnamefont
				{Houdr{\'{e}}}}, \bibinfo {author} {\bibfnamefont {R.~P.}\ \bibnamefont
				{Stanley}}, \ and\ \bibinfo {author} {\bibfnamefont {M.}~\bibnamefont
				{Ilegems}},\ }\href {\doibase 10.1103/PhysRevA.53.2711} {\bibfield  {journal}
			{\bibinfo  {journal} {Phys. Rev. A}\ }\textbf {\bibinfo {volume} {53}},\
			\bibinfo {pages} {2711} (\bibinfo {year} {1996})}\BibitemShut {NoStop}%
		\bibitem [{\citenamefont {Long}\ and\ \citenamefont
			{Simpkins}(2015)}]{Long2015}%
		\BibitemOpen
		\bibfield  {author} {\bibinfo {author} {\bibfnamefont {J.~P.}\ \bibnamefont
				{Long}}\ and\ \bibinfo {author} {\bibfnamefont {B.~S.}\ \bibnamefont
				{Simpkins}},\ }\href {\doibase 10.1021/ph5003347} {\bibfield  {journal}
			{\bibinfo  {journal} {ACS Photonics}\ }\textbf {\bibinfo {volume} {2}},\
			\bibinfo {pages} {130} (\bibinfo {year} {2015})}\BibitemShut {NoStop}%
		\bibitem [{\citenamefont {Meystre}\ and\ \citenamefont
			{Sargent}(2007)}]{Meystre2007}%
		\BibitemOpen
		\bibfield  {author} {\bibinfo {author} {\bibfnamefont {P.}~\bibnamefont
				{Meystre}}\ and\ \bibinfo {author} {\bibfnamefont {M.}~\bibnamefont
				{Sargent}},\ }\href@noop {} {\emph {\bibinfo {title} {{Elements of Quantum
						Optics}}}},\ \bibinfo {edition} {4th}\ ed.\ (\bibinfo  {publisher} {Springer
			Science \& Business Media},\ \bibinfo {address} {New York},\ \bibinfo {year}
		{2007})\BibitemShut {NoStop}%
		\bibitem [{\citenamefont {Lynden-Bell}\ and\ \citenamefont
			{McDonald}(1981)}]{Lynden-Bell1981}%
		\BibitemOpen
		\bibfield  {author} {\bibinfo {author} {\bibfnamefont {R.}~\bibnamefont
				{Lynden-Bell}}\ and\ \bibinfo {author} {\bibfnamefont {I.}~\bibnamefont
				{McDonald}},\ }\href {\doibase 10.1080/00268978100102181} {\bibfield
			{journal} {\bibinfo  {journal} {Mol. Phys.}\ }\textbf {\bibinfo {volume}
				{43}},\ \bibinfo {pages} {1429} (\bibinfo {year} {1981})}\BibitemShut
		{NoStop}%
		\bibitem [{\citenamefont {Impey}\ \emph {et~al.}(1982)\citenamefont {Impey},
			\citenamefont {Madden},\ and\ \citenamefont {McDonald}}]{Impey1982}%
		\BibitemOpen
		\bibfield  {author} {\bibinfo {author} {\bibfnamefont {R.}~\bibnamefont
				{Impey}}, \bibinfo {author} {\bibfnamefont {P.}~\bibnamefont {Madden}}, \
			and\ \bibinfo {author} {\bibfnamefont {I.}~\bibnamefont {McDonald}},\ }\href
		{\doibase 10.1080/00268978200101361} {\bibfield  {journal} {\bibinfo
				{journal} {Mol. Phys.}\ }\textbf {\bibinfo {volume} {46}},\ \bibinfo {pages}
			{513} (\bibinfo {year} {1982})}\BibitemShut {NoStop}%
		\bibitem [{\citenamefont {Miller}\ and\ \citenamefont
			{Manolopoulos}(2005)}]{Miller2005}%
		\BibitemOpen
		\bibfield  {author} {\bibinfo {author} {\bibfnamefont {T.~F.}\ \bibnamefont
				{Miller}}\ and\ \bibinfo {author} {\bibfnamefont {D.~E.}\ \bibnamefont
				{Manolopoulos}},\ }\href {\doibase 10.1063/1.2074967} {\bibfield  {journal}
			{\bibinfo  {journal} {J. Chem. Phys.}\ }\textbf {\bibinfo {volume} {123}},\
			\bibinfo {pages} {154504} (\bibinfo {year} {2005})}\BibitemShut {NoStop}%
		\bibitem [{\citenamefont {Kapil}\ \emph {et~al.}(2019)\citenamefont {Kapil},
			\citenamefont {Rossi}, \citenamefont {Marsalek}, \citenamefont {Petraglia},
			\citenamefont {Litman}, \citenamefont {Spura}, \citenamefont {Cheng},
			\citenamefont {Cuzzocrea}, \citenamefont {Mei{\ss}ner}, \citenamefont
			{Wilkins}, \citenamefont {Helfrecht}, \citenamefont {Juda}, \citenamefont
			{Bienvenue}, \citenamefont {Fang}, \citenamefont {Kessler}, \citenamefont
			{Poltavsky}, \citenamefont {Vandenbrande}, \citenamefont {Wieme},
			\citenamefont {Corminboeuf}, \citenamefont {K{\"{u}}hne}, \citenamefont
			{Manolopoulos}, \citenamefont {Markland}, \citenamefont {Richardson},
			\citenamefont {Tkatchenko}, \citenamefont {Tribello}, \citenamefont {{Van
					Speybroeck}},\ and\ \citenamefont {Ceriotti}}]{Kapil2019}%
		\BibitemOpen
		\bibfield  {author} {\bibinfo {author} {\bibfnamefont {V.}~\bibnamefont
				{Kapil}}, \bibinfo {author} {\bibfnamefont {M.}~\bibnamefont {Rossi}},
			\bibinfo {author} {\bibfnamefont {O.}~\bibnamefont {Marsalek}}, \bibinfo
			{author} {\bibfnamefont {R.}~\bibnamefont {Petraglia}}, \bibinfo {author}
			{\bibfnamefont {Y.}~\bibnamefont {Litman}}, \bibinfo {author} {\bibfnamefont
				{T.}~\bibnamefont {Spura}}, \bibinfo {author} {\bibfnamefont
				{B.}~\bibnamefont {Cheng}}, \bibinfo {author} {\bibfnamefont
				{A.}~\bibnamefont {Cuzzocrea}}, \bibinfo {author} {\bibfnamefont {R.~H.}\
				\bibnamefont {Mei{\ss}ner}}, \bibinfo {author} {\bibfnamefont {D.~M.}\
				\bibnamefont {Wilkins}}, \bibinfo {author} {\bibfnamefont {B.~A.}\
				\bibnamefont {Helfrecht}}, \bibinfo {author} {\bibfnamefont {P.}~\bibnamefont
				{Juda}}, \bibinfo {author} {\bibfnamefont {S.~P.}\ \bibnamefont {Bienvenue}},
			\bibinfo {author} {\bibfnamefont {W.}~\bibnamefont {Fang}}, \bibinfo {author}
			{\bibfnamefont {J.}~\bibnamefont {Kessler}}, \bibinfo {author} {\bibfnamefont
				{I.}~\bibnamefont {Poltavsky}}, \bibinfo {author} {\bibfnamefont
				{S.}~\bibnamefont {Vandenbrande}}, \bibinfo {author} {\bibfnamefont
				{J.}~\bibnamefont {Wieme}}, \bibinfo {author} {\bibfnamefont
				{C.}~\bibnamefont {Corminboeuf}}, \bibinfo {author} {\bibfnamefont {T.~D.}\
				\bibnamefont {K{\"{u}}hne}}, \bibinfo {author} {\bibfnamefont {D.~E.}\
				\bibnamefont {Manolopoulos}}, \bibinfo {author} {\bibfnamefont {T.~E.}\
				\bibnamefont {Markland}}, \bibinfo {author} {\bibfnamefont {J.~O.}\
				\bibnamefont {Richardson}}, \bibinfo {author} {\bibfnamefont
				{A.}~\bibnamefont {Tkatchenko}}, \bibinfo {author} {\bibfnamefont {G.~A.}\
				\bibnamefont {Tribello}}, \bibinfo {author} {\bibfnamefont {V.}~\bibnamefont
				{{Van Speybroeck}}}, \ and\ \bibinfo {author} {\bibfnamefont
				{M.}~\bibnamefont {Ceriotti}},\ }\href {\doibase 10.1016/j.cpc.2018.09.020}
		{\bibfield  {journal} {\bibinfo  {journal} {Comput. Phys. Commun.}\ }\textbf
			{\bibinfo {volume} {236}},\ \bibinfo {pages} {214} (\bibinfo {year}
			{2019})}\BibitemShut {NoStop}%
		\bibitem [{\citenamefont {Li}(2020)}]{TELi2020Github}%
		\BibitemOpen
		\bibfield  {author} {\bibinfo {author} {\bibfnamefont {T.~E.}\ \bibnamefont
				{Li}},\ }\href {https://github.com/TaoELi/cavity-md-ipi} {}\bibinfo
		{howpublished} {https://github.com/TaoELi/cavity-md-ipi} (\bibinfo {year}
		{2020})\BibitemShut {NoStop}%
	\end{thebibliography}

	%
	
\end{document}


\title{Supporting Information for "Cavity Molecular Dynamics Simulations of Liquid Water under Vibrational  Ultrastrong Coupling"}
	
	\author{Tao E. Li}%
	\affiliation{Department of Chemistry, University of Pennsylvania, Philadelphia, Pennsylvania 19104, USA}
	
	\author{Joseph E. Subotnik}
	\affiliation{Department of Chemistry, University of Pennsylvania, Philadelphia, Pennsylvania 19104, USA}
	
	\author{Abraham Nitzan} 
	\email{anitzan@sas.upenn.edu}
	\affiliation{Department of Chemistry, University of Pennsylvania, Philadelphia, Pennsylvania 19104, USA}
	\affiliation{School of Chemistry, Tel Aviv University, Tel Aviv 69978, Israel}

	\maketitle

	\setcounter{equation}{0}
	\setcounter{figure}{0}
	\setcounter{table}{0}
	\renewcommand{\theequation}{S\arabic{equation}}
	\renewcommand{\thefigure}{S\arabic{figure}}
	\renewcommand{\bibnumfmt}[1]{[S#1]}
	\renewcommand{\citenumfont}[1]{S#1}

		\section{Details on Classical Molecular Dynamics}\label{sec:MD}
	The quantum Hamiltonian in Eq. (3), although depending only on the nuclear and photonic degrees of freedom, is still too expensive to evolve exactly. The simplest approximation we can make is the classical approximation, i.e., all quantum operators are mapped to the corresponding classical observables, which leads to the following classical Hamiltonian:
	\begin{subequations}\label{eq:H_MD}
		\begin{align}
		H_\text{QED}^{\text{G}} &= H_\text{M}^{\text{G}} + H_\text{F}^{\text{G}} \\
		H_\text{M}^{\text{G}} &=  \sum_{n=1}^{N}\left(\sum_{j\in n} \frac{\mathbf{P}_{nj}^2}{2 M_{nj}} + V^{(n)}_{g}(\{\mathbf{R}_{nj}\})\right) + 
		\sum_{n=1}^{N}\sum_{l>n}
		V_{\text{inter}}^{(nl)}\\
		H_\text{F}^{\text{G}} &= \sum_{k,\lambda}
		\frac{\widetilde{p}_{k, \lambda}^2}{2 m_{k, \lambda}}  + 
		\frac{1}{2} m_{k,\lambda}\omega_{k,\lambda}^2 \left(
		\widetilde{q}_{k,\lambda} + \sum_{n=1}^{N} \frac{d_{ng, \lambda}}{\omega_{k,\lambda}\sqrt{\Omega\epsilon_0 m_{k,\lambda}}} 
		\right)^2 
		\end{align}
	\end{subequations}
	Eq. \eqref{eq:H_MD} serves as the starting point of this work. We note that one can go beyond the treatment here by propagating the quantum Hamiltonian (2) using  the path-integral technique  \cite{Tuckerman2010,Markland2018} and evolve the ring polymer Hamiltonian with $n$ copies of coupled classical trajectories (aka $n$ beads). 
	In the present manuscript, we focus on the classical system, deferring the path-integral calculation to a later study.
	
	In our classical MD simulations, the simulated system is represented by particles that obey the Newtonian equations of motion:
	\begin{subequations}\label{eq:EOM_MD}
		\begin{align}
		M_{nj}\ddot{\mathbf{R}}_{nj} &= \mathbf{F}_{nj}^{(0)} - 
		\sum_{k,\lambda}
		\left(\varepsilon_{k,\lambda} \widetilde{q}_{k,\lambda}
		+ \frac{\varepsilon_{k,\lambda}^2}{m_{k,\lambda} \omega_{k,\lambda}^2} \sum_{l=1}^{N} d_{lg,\lambda}
		\right)
		\frac{\partial d_{ng, \lambda}}{\partial \mathbf{R}_{nj}}
		\\
		m_{k,\lambda}\ddot{\widetilde{q}}_{k,\lambda} &= - m_{k,\lambda}\omega_{k,\lambda}^2 \widetilde{q}_{k,\lambda}
		-\varepsilon_{k,\lambda} \sum_{n=1}^{N}d_{ng,\lambda} \label{eq:EOM_MD-2}
		\end{align}
	\end{subequations}
	where the cavity-free force $\mathbf{F}_{nj}^{(0)}$ is calculated by $\mathbf{F}_{nj}^{(0)} = - \partial V_g^{(n)} / \partial \mathbf{R}_{nj} - \sum_{l\neq n} \partial V_{\text{inter}}^{(nl)}/\partial \mathbf{R}_{nj}$, and the
	coupling between particles representing photons and nuclear degrees of freedom is given by $\varepsilon_{k,\lambda} \equiv \sqrt{m_{k,\lambda} \omega_{k,\lambda}^2/\Omega \epsilon_0}$. 
	
	\subsection*{Periodic Boundary Condition}
	A realistic simulation for VSC or V-USC that corresponds to current observations requires a macroscopic number (say, $10^9 \sim 10^{11}$)  of molecules  \cite{Thomas2016,Lather2019,Hiura2018,Thomas2019_science}, which is far beyond our computational power if we simulate Eq. \eqref{eq:EOM_MD} directly. To proceed, we assume that the whole molecular ensemble can be divided into $N_{\text{cell}}$ periodic cells, in which the molecules evolve identically, i.e., we can approximate
	the second term on the right of Eq. \eqref{eq:EOM_MD-2} by $\sum_{n=1}^{N}d_{ng,\lambda} = N_{\text{cell}}\sum_{n=1}^{N_{\text{sub}}}d_{ng,\lambda}$, where $N_{\text{sub}} = N/N_{\text{cell}}$ denotes the number of molecules in a single cell. By further denoting $\dbtilde{q}_{k,\lambda} = \widetilde{q}_{k,\lambda} / \sqrt{N_{\text{cell}}}$, $\widetilde{\varepsilon}_{k,\lambda} = \sqrt{N_{\text{cell}}} \varepsilon_{k,\lambda}$, we can rewrite the equations of motion in Eq. \eqref{eq:EOM_MD} in a symmetric form
	\begin{subequations}\label{eq:EOM_MD_PBC_SI}
		\begin{align}
		M_{nj}\ddot{\mathbf{R}}_{nj} &= \mathbf{F}_{nj}^{(0)} - 
		\sum_{k,\lambda}
		\left(\widetilde{\varepsilon}_{k,\lambda} \dbtilde{q}_{k,\lambda}
		+ \frac{\widetilde{\varepsilon}_{k,\lambda}^2}{m_{k,\lambda} \omega_{k,\lambda}^2} \sum_{l=1}^{N_{\text{sub}}} d_{lg,\lambda}
		\right)
		\frac{\partial d_{ng, \lambda}}{\partial \mathbf{R}_{nj}} \label{eq:EOM_MD_PBC_SI-1}
		\\
		m_{k,\lambda}\ddot{\dbtilde{q}}_{k,\lambda} &= - m_{k,\lambda}\omega_{k,\lambda}^2 \dbtilde{q}_{k,\lambda}
		-\widetilde{\varepsilon}_{k,\lambda} \sum_{n=1}^{N_{\text{sub}}}d_{ng,\lambda}
		\end{align}
	\end{subequations}
	The form of Eq. \eqref{eq:EOM_MD_PBC_SI} has several advantages. First, we simulate the VSC of a macroscopic number of molecules by evolving molecules in a single cell plus the few photon modes that we are interested in. Second, when considering the dependence of Rabi splitting on molecular numbers, we can fix the number of molecules in a single cell ($N_{\text{sub}}$)  and vary only the coupling constant $\widetilde{\varepsilon}_{k,\lambda} = \sqrt{N_{\text{cell}}} \varepsilon_{k,\lambda}$. Such a change is very easy to implement in practice and has the physical interpretation of increasing the number of cells while leaving the number of molecules per cell and the size of the simulation cell fixed.

	\subsection*{q-TIP4P/F Water Force Field}
	The  question remains as to exactly how we will calculate the ground-state quantities $\mathbf{F}_{nj}^{(0)}$, $d_{ng,\lambda}$, and $\partial d_{ng, \lambda}/\partial \mathbf{R}_{nj}$. In general, these properties can be calculated by classical empirical force field or \textit{ab initio} electronic structure theory. For this initial work, we  use an empirical classical force field --- the q-TIP4P/F water model  \cite{Habershon2009} --- which provides a simple yet reliable description of both the equilibrium and dynamic properties of liquid water. 
	
	In the q-TIP4P/F model, the  pairwise intermolecular potential is characterized  by the Lennard-Jones potential between oxygen atoms plus the Coulombic interactions between partial changes:
	\begin{equation}
	V_{\text{inter}}^{(nl)} =  4\epsilon \left[\left(\frac{\sigma}{R^{\ch{OO}}_{nl}}\right)^{12} - \left(\frac{\sigma}{R^{\ch{OO}}_{nl}}\right)^{6} \right]
	+
	\sum_{i\in n} \sum_{j \in l} \frac{Q_i Q_j}{R_{ij}}
	\end{equation}
	where $R^{\ch{OO}}_{nl}$ denotes the distance between the oxygen atoms and $R_{ij}$ ($i \in n$ and $j \in l$) denotes the distance between the partial charge sites in molecules $n$ and $l$. Within a single \ch{H2O} molecule, two positive partial charge with magnitude $Q_{\ch{M}}/2$ are assigned to the hydrogen atoms, and the negatively charge site with magnitude $-Q_{\ch{M}}$ is placed at $\mathbf{R}_{\ch{M}}$:
	\begin{equation}
	\mathbf{R}_{\ch{M}} = \gamma \mathbf{R}_{\ch{O}} + \frac{1 - \gamma}{2} 
	\left(\mathbf{R}_{\ch{H1}} + \mathbf{R}_{\ch{H2}}\right)
	\end{equation}
	For parameters, $\epsilon = 0.1852 \text{\ kcal mol}^{-1}$, $\sigma = 3.1589 \ \si{\angstrom}$, $Q_{\ch{M}} = 1.1128\ |e|$ (where $e$ denotes the charge of the electron), and $\gamma = 0.73612$.

	The intramolecular interaction is characterized by
	\begin{equation}
	V_{\text{g}}^{(n)} =  V_{\ch{OH}}(R_{n1}) + V_{\ch{OH}}(R_{n1}) 
	+ \frac{1}{2}k_{\theta} \left(\theta_n - \theta_{\text{eq}}\right)^2
	\end{equation}
	where
	\begin{equation}
	V_{\ch{OH}}(r) = D_r \left[\alpha_r^2 \left(r -r_{\text{eq}}\right)^2
	- \alpha_r^3 \left(r -r_{\text{eq}}\right)^3
	+ \frac{7}{12}\alpha_r^4 \left(r -r_{\text{eq}}\right)^4
	\right]
	\end{equation}
	Here, $R_{n1}$ and $R_{n2}$ denote the lengths of two \ch{O-H} bonds, $\theta_n$ 
	and $\theta_\text{eq}$ denote the \ch{H-O-H} angle and the equilibrium angle. For parameters, $D_r = 116.09 \text{\ kcal mol}^{-1}$, $\alpha_r = 2.287 \si{\angstrom}^{-1}$, $r_{\text{eq}} = 0.9419 \si{\angstrom}$, $k_{\theta} = 87.85 \text{ \ kcal mol}^{-1}\text{rad}^{-2}$, and $\theta_{\text{eq}} = 107.4 \text{ deg}$.
	
	Given the q-TIP4P/F force field, one can  easily calculate the cavity-free force $\mathbf{F}_{nj}^{(0)}$ as a function of the nuclear configurations by standard molecular dynamics packages. The dipole moment is given by
	\begin{equation}
	\begin{aligned}
	d_{ng, \lambda} &=\left[ \frac{Q_{\ch{M}}}{2}\left(\mathbf{R}_{n\ch{H1}}
	+ \mathbf{R}_{n\ch{H2}}\right) - Q_{\ch{M}}\mathbf{R}_{n\ch{M}}\right] \cdot \vxi_{\lambda}\\
	&= \left[ \frac{\gamma Q_{\ch{M}}}{2}
	\left(\mathbf{R}_{n\ch{H1}}
	+ \mathbf{R}_{n\ch{H2}}\right) 
	- \gamma Q_{\ch{M}} \mathbf{R}_{n\ch{O}}\right]\cdot \vxi_{\lambda}
	\end{aligned}
	\end{equation}
	and the derivative $\partial d_{ng,\lambda}/\partial \mathbf{R}_{nj}$ is straightforward. In calculating the IR spectrum, the total dipole moment $\vmu_S$ is given by $\vmu_S \cdot \vxi_{\lambda} = \sum_{n=1}^{N_{\text{sub}}} d_{ng,\lambda}$.
	
	\subsection*{Implementation Details}
	
	\begin{figure}
		\centering
		\includegraphics[width=0.5\linewidth]{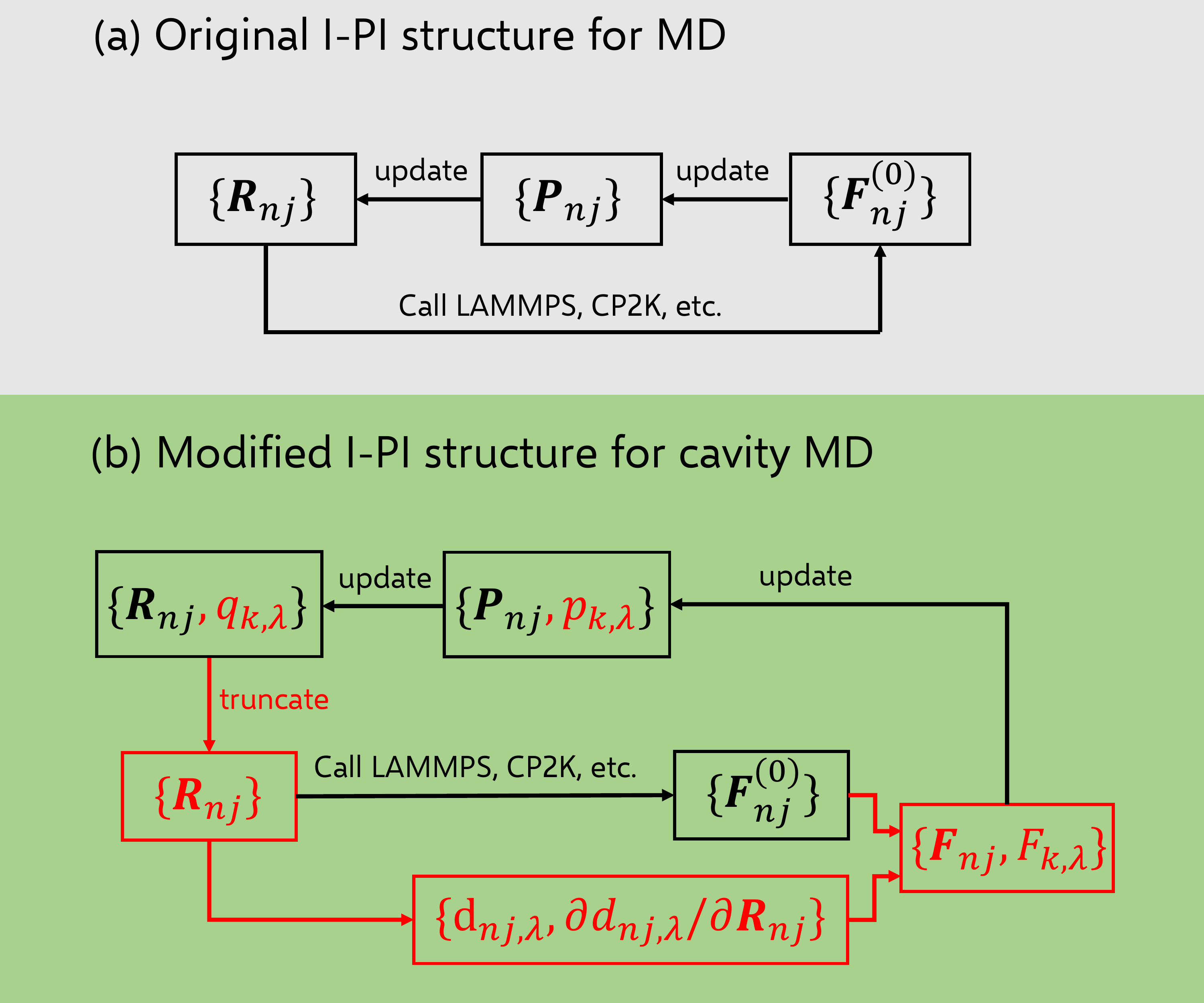}
		\caption{Illustration of  the algorithm structures of (a) the original I-PI for MD simulations and (b) our modified I-PI structure for cavity MD simulations, where the modification is labeled in red.}
		\label{fig:demo}
	\end{figure}
	
	We have implemented the above cavity MD scheme by modifying an open-source MD package I-PI  \cite{Kapil2019}, which was designed for both classical and path-integral MD simulations. The general structure of I-PI is illustrated as the gray region in Fig. \ref{fig:demo}: At every time step, given the molecular positions $\{\mathbf{R}_{nj}\}$, the forces  $\{\mathbf{F}_{nj}^{(0)}\}$  were calculated by calling the external package LAMMPS (for classical MD)  \cite{Plimpton1995}. The package CP2K \cite{Hutter2014} could be used for \textit{ab initio} MD. After calculating the forces, the momenta $\{\mathbf{P}_{nj}\}$ and positions $\{\mathbf{R}_{nj}\}$ are updated accordingly.
	
	Our modification is illustrated as the green region in Fig. \ref{fig:demo}. We store both the nuclear and photonic degrees of freedom in I-PI. At every time step, we first truncate a nuclear position array $\{\mathbf{R}_{nj}\}$ from the total (nuclear + photonic) position array, and then use the interface of I-PI to calculate the cavity-free forces $\{\mathbf{F}_{nj}^{(0)}\}$. We also calculate the dipole moments and their derivatives from the nuclear position array $\{\mathbf{R}_{nj}\}$. With the cavity-free forces and the dipole moments, we calculate the overall forces on all nuclei and photons $\{\mathbf{F}_{nj}, F_{k,\lambda}\}$ (the right hand side of Eq. \eqref{eq:EOM_MD_PBC_SI}).  After calculating the forces, we use the interface of I-PI to update momenta and positions. 
	
	Due to the user-friendly structure of I-PI, the current cavity MD code should be easily generalized to the cases of \textit{ab initio} calculation and path-integral cavity MD simulations, results which will be reported in a separate publication.

	\subsection*{Simulation Details}\label{sec:simu_details}
	
	\begin{figure}
		\centering
		\includegraphics[width=0.5\linewidth]{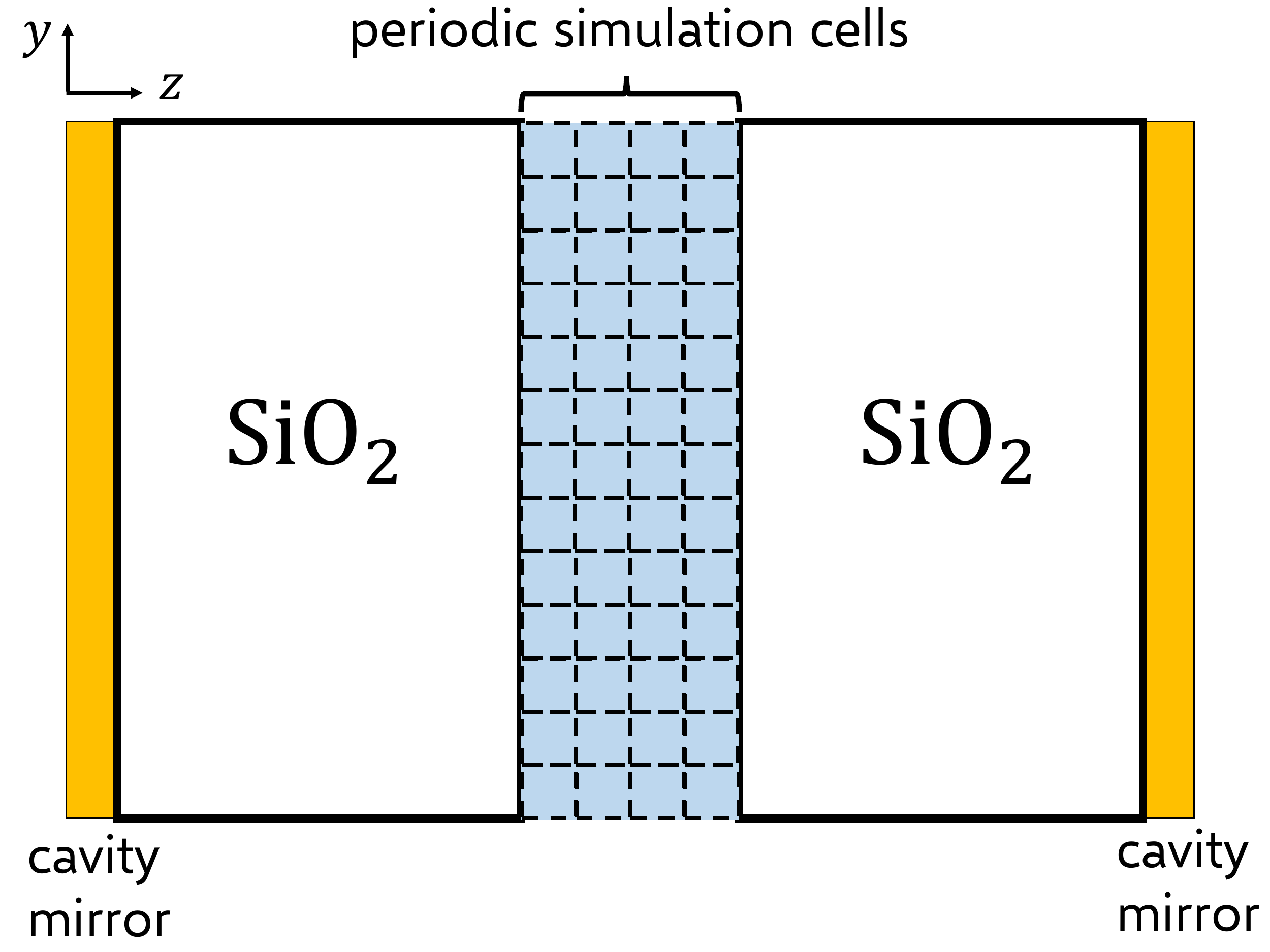}
		\caption{The structure of the cavity for our simulation. Water molecules are constrained at the center of the cavity by a pair of thick \ch{SiO2} layers.}
		\label{fig:cavity_structure}
	\end{figure}
	
	We consider the following scenario for simulation. As shown in Fig. \ref{fig:cavity_structure}, the cavity is placed along the $z$-axis. A pair of thick \ch{SiO2}  layers are placed between the cavity mirrors so that the water molecules can move  freely only in a small region (but still on the order of microns) near the cavity center. Such additional \ch{SiO2} layers are used (i) to ensure the intermolcular interactions between \ch{H2O} molecules are the same as those in free space, and (ii) to validate the long-wave approximation that we have taken from the very beginning. We consider only two cavity modes with polarization directions $\vxi_{\lambda}$ along $x$ and $y$ directions, both of which are resonant with the \ch{O-H} stretch mode. We set the auxiliary mass for the two photons as $m_{k, \lambda} = 1$ a.u. (atomic units).

	\section{Simplified 1D Model for V-USC}\label{sec:1d}
	
	Starting from the classical Hamiltonian for V-USC (Eq. \eqref{eq:H_MD}), let us assume that (i) the molecules are non-interacting 1D harmonic oscillators, (ii) the dipole moment for a single molecule is linear (i.e., $d_{ng,\lambda} = d_0 x_n$), and (iii) only a single cavity mode is considered. With these simplifications, the Hamiltonian can be written as:
	\begin{subequations}\label{eq:H_1D}
		\begin{equation}
		H_{\text{QED}}^{\text{G}} = \sum_{n=1}^{N}\frac{p_n^2}{2} + \frac{p_c^2}{2} + V(\{x_n\}, x_c)
		\end{equation}
		where 
		\begin{equation}
		V(\{x_n, x_c\}) = \sum_{n=1}^{N} \frac{1}{2}\omega_0^2 x_n^2 + \frac{1}{2}\omega_c^2\left(x_c + \frac{2g_0}{\omega_c} \sum_{n=1}^{N} x_n \right)^2
		\end{equation}
	\end{subequations}
	Here, $g_0 \equiv d_0 / 2\sqrt{\Omega \epsilon_0}$, and we have assumed all masses to be $1$. Note that the self-dipole term (the $\left(\sum_{n=1}^{N} x_n\right)^2$ term in the expanded square above) is necessary for studying V-USC.
	With the Hamiltonian in Eq. \eqref{eq:H_1D},
	the equations of motion now read
	\begin{subequations}
		\begin{align}
		\ddot{x}_n &= - \omega_0^2 x_n - 2 g_0 \omega_c x_c - 4 g_0^2 \sum_{l=1}^{N} x_l \\
		\ddot{x}_c &= - \omega_c^2 x_c - 2 g_0 \omega_c \sum_{n=1}^{N} x_n \label{eq:EOM_1d_2}
		\end{align}
	\end{subequations}
	Let us define the bright mode as $x_B = \frac{1}{\sqrt{N}} \sum_{n=1}^{N} x_n$, so that the equations of motion for the bright mode and the cavity mode become
	\begin{subequations}\label{eq:EOM_1d}
		\begin{align}
		\ddot{x}_B &= - \omega_0^2 x_B - \omega_c \Omega_N x_c - \Omega_N^2 x_B \\
		\ddot{x}_c &= - \omega_c^2 x_c - \omega_c\Omega_N x_B 
		\end{align}
	\end{subequations}
	where $\Omega_N = 2\sqrt{N} g_0$ is the usual Rabi frequency. In the matrix form, the above equations can be written as
	\begin{equation}
	\ddot{\vec{x}} = - K \vec{x}
	\end{equation}
	where $\vec{x} = (x_B, x_c)^{T}$ and 
	\begin{equation}\label{eq:K}
	K=
	\begin{pmatrix}
	\omega_0^2 + \Omega_N^2 & \omega_c \Omega_N  \\
	\omega_c \Omega_N & \omega_c^2
	\end{pmatrix}
	\end{equation}
	Note that the $\Omega_N^2$ term above comes from the self-dipole term.
	
	\subsection*{Polariton frequencies}
	The polariton frequencies ($\omega_{\pm}$) can be determined by solving the eigenvalues of the matrix $K$:
	\begin{equation}
	\omega_{\pm}^2 = \frac{1}{2}\left[\omega_0^2 + \Omega_N^2 + \omega_c^2 \pm \sqrt{(\omega_0^2 + \Omega_N^2 + \omega_c^2)^2 - 4\omega_0^2\omega_c^2}\right]
	\end{equation}
	At resonance ($\omega_c = \omega_0$), the polariton frequencies are reduced to
	\begin{equation}\label{eq:polariton_freq_resonance}
	\omega_{\pm}^2 = \omega_0^2 + \frac{\Omega_N^2}{2} \pm \Omega_N\sqrt{\omega_0^2 + \frac{\Omega_N^2}{4}}
	\end{equation}
	In the VSC limit ($\Omega_N \ll \omega_0$), Eq. \eqref{eq:polariton_freq_resonance} can be further simplified as
	\begin{equation}
	\omega_{\pm} \approx \sqrt{\omega_0^2 \pm \Omega_N \omega_0} \approx \omega_0 \pm \frac{\Omega_N}{2}
	\end{equation}
	which is the usual strong-coupling result.
	
	\subsection*{IR spectrum}
	The IR spectrum of molecules is calculated by Eq. (5). With our 1D model, the IR spectrum is expressed as
	\begin{equation}\label{eq:IR_1d}
	n(\omega)\alpha(\omega) \propto \omega^2 \int_{-\infty}^{+\infty} e^{-i\omega t}\avg{x_B(0) x_B(t)} dt
	\end{equation}
	where we have neglected all prefactors (including the temperature as we take room temperature throughout this manuscript). According to Eq. \eqref{eq:EOM_1d}, the solution of $x_B(t)$ is 
	\begin{subequations}\label{eq:xB_1d}
		\begin{equation}
	\begin{aligned}
		x_B(t) &= \left[x_B(0)\cos^2\left(\frac{\theta}{2}\right) + x_c(0)\cos\left(\frac{\theta}{2}\right)\sin\left(\frac{\theta}{2}\right)\right] e^{i\omega_{+}t} \\
		&
	+ \left[x_B(0)\sin^2\left(\frac{\theta}{2}\right) - x_c(0)\cos\left(\frac{\theta}{2}\right)\sin\left(\frac{\theta}{2}\right)\right] e^{i\omega_{-}t}
	\end{aligned}
	\end{equation}
	where 
	\begin{equation}\label{eq:theta}
	\tan\left(\theta\right) = 2\omega_c \Omega_N / \left(\omega_0^2 + \Omega_N^2 - \omega_c^2\right)
	\end{equation}
	\end{subequations}
	By substituting Eq. \eqref{eq:xB_1d} into Eq. \eqref{eq:IR_1d} and using $\avg{x_B(0) x_c(0)} = 0$, we obtain  
	\begin{equation}
	n(\omega)\alpha(\omega) \propto \omega^2 \left[\cos^2\left(\frac{\theta}{2}\right) \delta(\omega - \omega_{+}) + \sin^2\left(\frac{\theta}{2}\right) \delta(\omega - \omega_{-})\right]
	\end{equation}
	The integrated peak areas for LP and UP are 
	\begin{subequations}
		\begin{align}
		I_{\text{LP}} &\propto \omega_{-}^2 \sin^2\left(\frac{\theta}{2}\right) \\
		I_{\text{UP}} &\propto \omega_{+}^2 \cos^2\left(\frac{\theta}{2}\right)
		\end{align}
	\end{subequations}
	From the above, we find that the asymmetric peaks come from two origins: (i) the prefactor $\omega_{\pm}^2$ and (ii) the self-dipole term in the dipole-gauge Hamiltonian. While the first origin is trivial, the second origin can be understood as follows.
	If we had neglected the self-dipole term, we would naively take $\omega_0^2 + \Omega_N^2 \rightarrow \omega_0^2$ in Eq. \eqref{eq:K} and obtain a different expression for $\theta$, $\tan\left(\theta\right) = 2\omega_c \Omega_N / \left(\omega_0^2 - \omega_c^2\right)$ (where the $\Omega_N^2$ term now vanishes compared to Eq. \eqref{eq:theta}).
	At resonance, we would obtain that $\tan\left(\theta\right)=\infty$, i.e., $\theta = \pi / 2$ and $\cos^2\left(\theta/2\right) = \sin^2\left(\theta/2\right)= 1/2$. However, because of the $\Omega_N^2$ term, $\theta < \pi/2 $, which implies $\sin^2\left(\theta/2\right) < \cos^2\left(\theta/2\right)$. 
	In other words, when $\omega_c = \omega_0$, the self-dipole term forces the LP to be further suppressed and the UP be further enhanced.

	\section{Velocity Autocorrelation Function of \ch{O-H} Bond}\label{sec:vacf-OH}
	\begin{figure}
		\centering
		\includegraphics[width=0.5\linewidth]{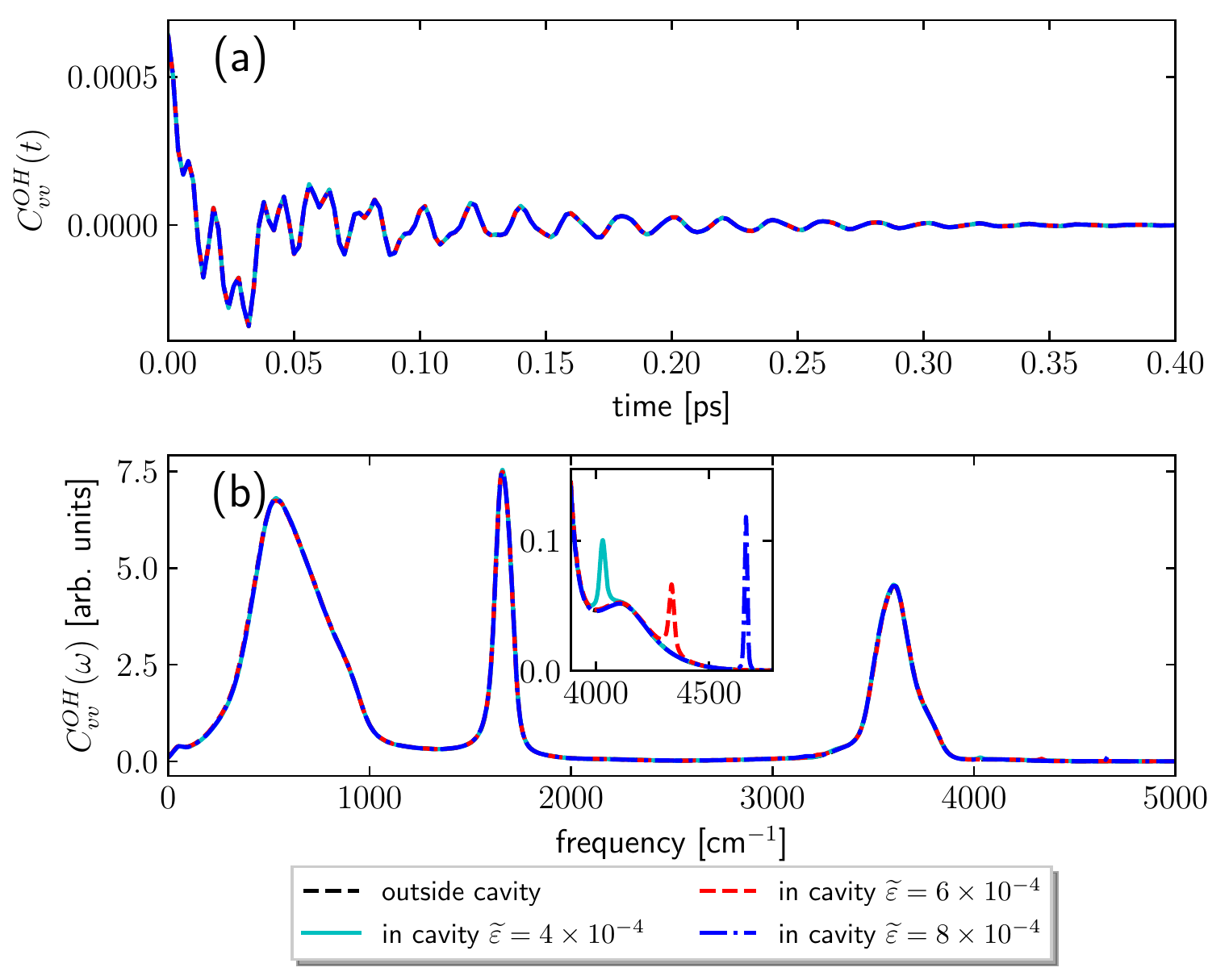}
		\caption{VACF of \ch{O-H} bond for individual \ch{H2O} molecules plotted in the same manner as Fig. 7. Note that under V-USC, a small side peak also emerges in the spectrum which corresponds to the UP frequency (see the zoom-in insert of Fig. \ref{fig:vacf-OH}b).}
		\label{fig:vacf-OH}
	\end{figure}
	
	We report the VACF of the \ch{O-H} bond for individual \ch{H2O} molecules in Fig. \ref{fig:vacf-OH}, which is plotted in the same manner as Fig. 7. While the VACF of \ch{O-H} is largely the same for outside (black dashed line) or inside (lines with color) the cavity, we also find a small side peak in the spectrum which corresponds to the UP frequency; see the zoom-in insert of Fig. \ref{fig:vacf-OH}b. However, compared with Fig. \ref{fig:vacf-OH}b, the side peak is much less intense.
	
	\section{Multimode Rabi Splitting}\label{sec:multimode}
	The results presented in the manuscript and above are limited to the case of a single-mode cavity. In this section, we consider the case when a multimode cavity is coupled to liquid water.
	In detail, we consider a cavity of $N_m$ ($N_m > 1$) different cavity modes with frequencies $\omega_{k,\lambda} = m\pi c/L_c(N_m)$ ($m = 1, 2, \cdots, N_m$), where $L_c(N_m)$ denotes the cavity length which depends on $N_m$. To best isolate and analyze the effect of the multimode cavity, we set the middle cavity mode ($\omega_{k,\lambda} = N_m\pi c/2L_c(N_m)$) at resonance with the \ch{O-H} stretch mode (with frequency $\omega_0 = 3550 \text{\ cm}^{-1}$) , i.e., the cavity length $L_c(N_m)$ is increased when more cavity modes are considered.
	Note that when increasing the cavity length, both the cavity volume $\Omega$ and the number of simulation cells $N_{\text{cell}}$ also increase and with the same proportion. As such, the effective light-matter coupling strength  $\widetilde{\varepsilon}_{k,\lambda} = \sqrt{N_{\text{cell}}}\varepsilon_{k,\lambda}
	= \sqrt{ N_{\text{cell}} m_{k,\lambda} \omega_{k,\lambda}^2 / \Omega \epsilon_0}$ is kept the same for the middle cavity mode ($\omega_{k,\lambda} = N_m\pi c/2L_c(N_m) = \omega_0$) and one should expect the Rabi splitting for the \ch{O-H} stretch mode to remain the same as well. Note that, just as for the case of the single-mode cavity, each cavity mode contains two polarization directions ($x$- and $y$-direction).

	\begin{figure}
		\centering
		\includegraphics[width=0.5\linewidth]{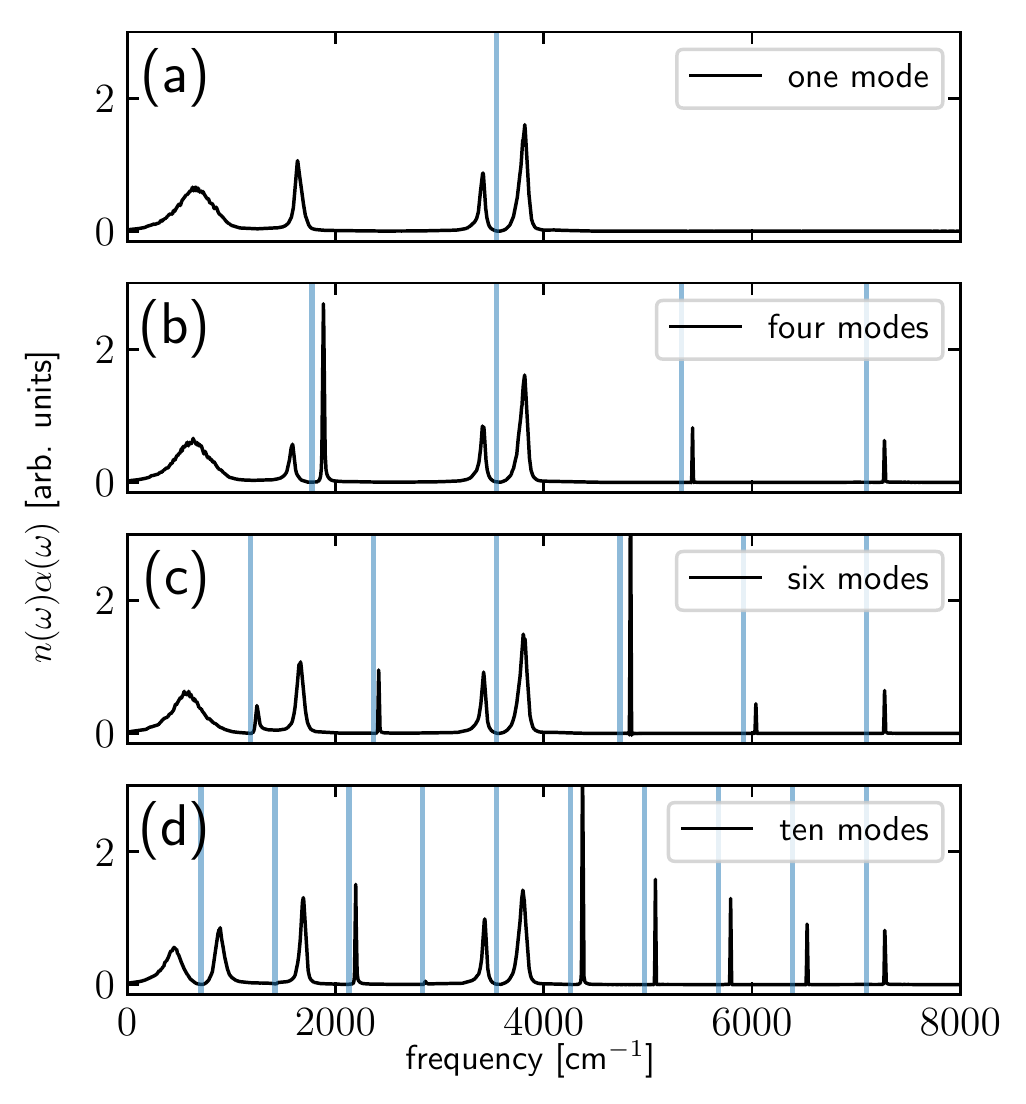}
		\caption{Simulated IR spectrum of liquid water in a multimode cavity. From top to bottom, we plot the results when a (a) single-mode, (b) four-mode, (c) six-mode, and (d) ten-mode cavity is coupled to liquid water, respectively. The vertical blue lines denote the frequencies of the cavity modes.  The effective coupling strength $\widetilde{\varepsilon}$ is always set as $2\times 10^{-4}$ a.u. for the middle cavity mode which is resonantly coupled to the \ch{O-H} stretch mode ($\sim 3550 \text{\ cm}^{-1}$). All other simulation details are the same as Fig. 1. Note that apart from the Rabi splitting for the \ch{O-H} stretch mode, Fig. \ref{fig:IR_multimode}b (or \ref{fig:IR_multimode}d) also shows the Rabi splitting between the fundamental cavity mode and the \ch{H-O-H} bending band near $1650 \text{\ cm}^{-1}$ (or sometimes the intermolecular librational vibration near $700 \text{\ cm}^{-1}$). Note that the decoupled cavity modes can also be found in the IR spectrum; however the observed frequencies are slightly larger than those of the cavity modes due to the self-dipole contribution; see Eq. \eqref{eq:EOM_MD_PBC_SI-1}.}
		\label{fig:IR_multimode}
	\end{figure}

	Fig. \ref{fig:IR_multimode}a-d plots the simulated IR spectrum of liquid water coupled to a (a) single-mode, (b) four-mode, (c) six-mode, or (d) ten-mode cavity. The blue lines denote the frequencies of the included cavity modes. 
	Here, the effective light-matter coupling strength for the middle cavity mode is always set as $\widetilde{\varepsilon} = 2\times 10^{-4} \text{\ cm}^{-1}$, and as one might expect from the argument above, indeed we do the same Rabi splitting for the \ch{O-H} stretch mode ($\sim 3400 \text{\ cm}^{-1}$) for each multimode case. Very interestingly, in Fig. \ref{fig:IR_multimode}b (or Fig. \ref{fig:IR_multimode}d), the fundamental cavity mode is  resonantly coupled to the \ch{H-O-H} bending band near $1650 \text{\ cm}^{-1}$ (or intermolecular librational vibration near $700 \text{\ cm}^{-1}$) and an additional Rabi splitting is also observed. As shown in the figure, apart from the Rabi splittings, the decoupled cavity modes can also be found in the IR spectrum. However, the observed frequencies are slightly larger than the cavity modes due to the self-dipole contribution; see Eq. \eqref{eq:EOM_MD_PBC_SI-1}. Notably, the intensities of these decoupled cavity peaks can vary significantly.
	
		\begin{figure}
		\centering
		\includegraphics[width=0.5\linewidth]{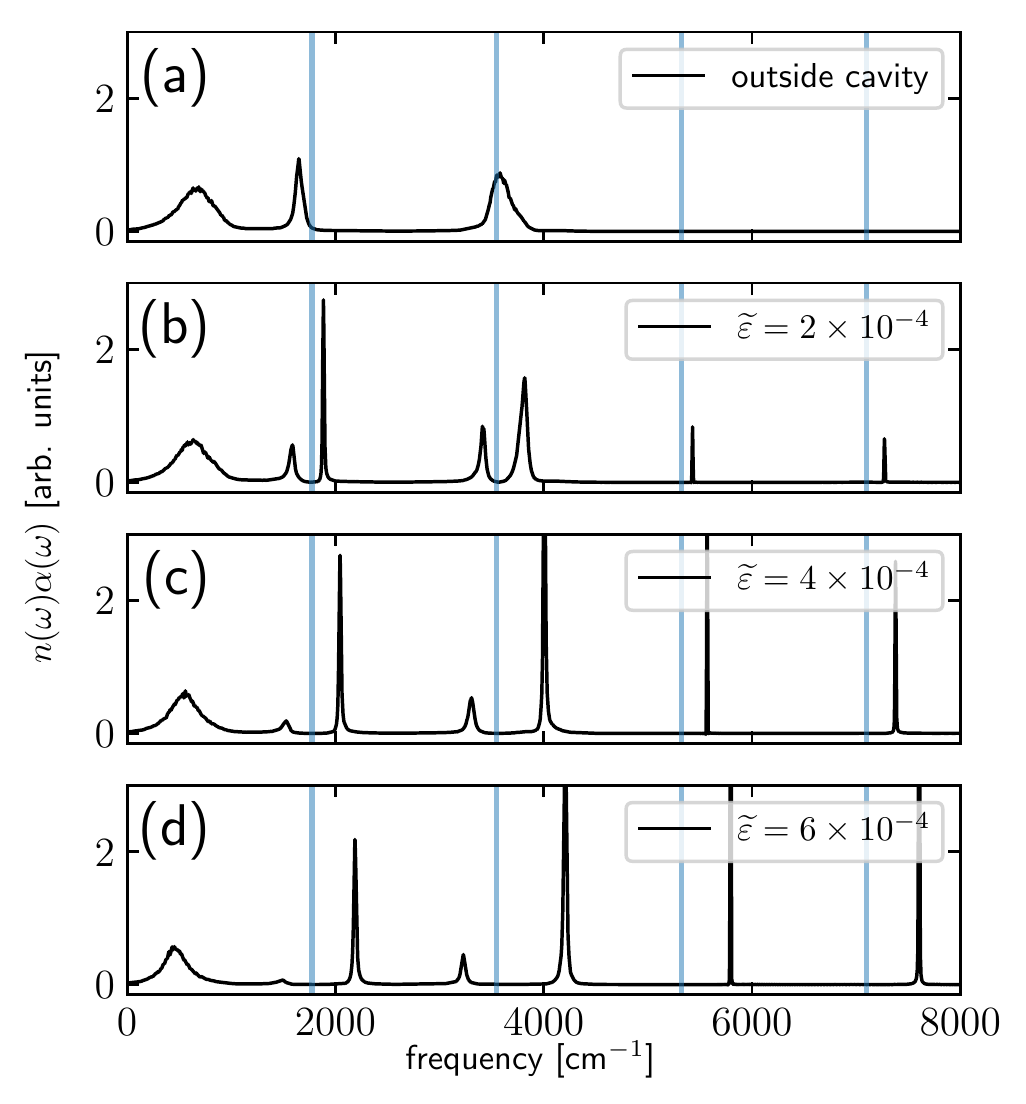}
		\caption{Simulated IR spectrum of liquid water in a four-mode cavity either (a) outside the cavity; or inside the cavity with effective  coupling strength $\widetilde{\varepsilon}$ for the middle cavity mode set as (b) $2\times 10^{-4}$, (c) $4\times 10^{-4}$, and (d) $6\times 10^{-4}$ a.u.. The vertical blue lines denote the frequencies of the cavity modes. All simulation details are the same as Fig. \ref{fig:IR_multimode}.}
		\label{fig:IR_multimode_g0}
	\end{figure}

	Fig. \ref{fig:IR_multimode_g0} plots the IR spectrum of liquid water in the four-mode cavity with different coupling strengths. Compared with the IR spectrum outside the cavity (Fig. \ref{fig:IR_multimode_g0}a),  in Fig. \ref{fig:IR_multimode_g0}b-d, increased Rabi splittings for both the \ch{O-H} stretch mode ($\sim 3400 \text{\ cm}^{-1}$) and the \ch{H-O-H} bending band  ($\sim 1650 \text{\ cm}^{-1}$) are observed when the effective coupling strength $\widetilde{\varepsilon}$ for the middle photon mode (which is resonantly coupled to the \ch{O-H} stretch mode) is increased from $2\times 10^{-4}$ to $6\times 10^{-4}$ a.u..

		\begin{figure}
		\centering
		\includegraphics[width=0.5\linewidth]{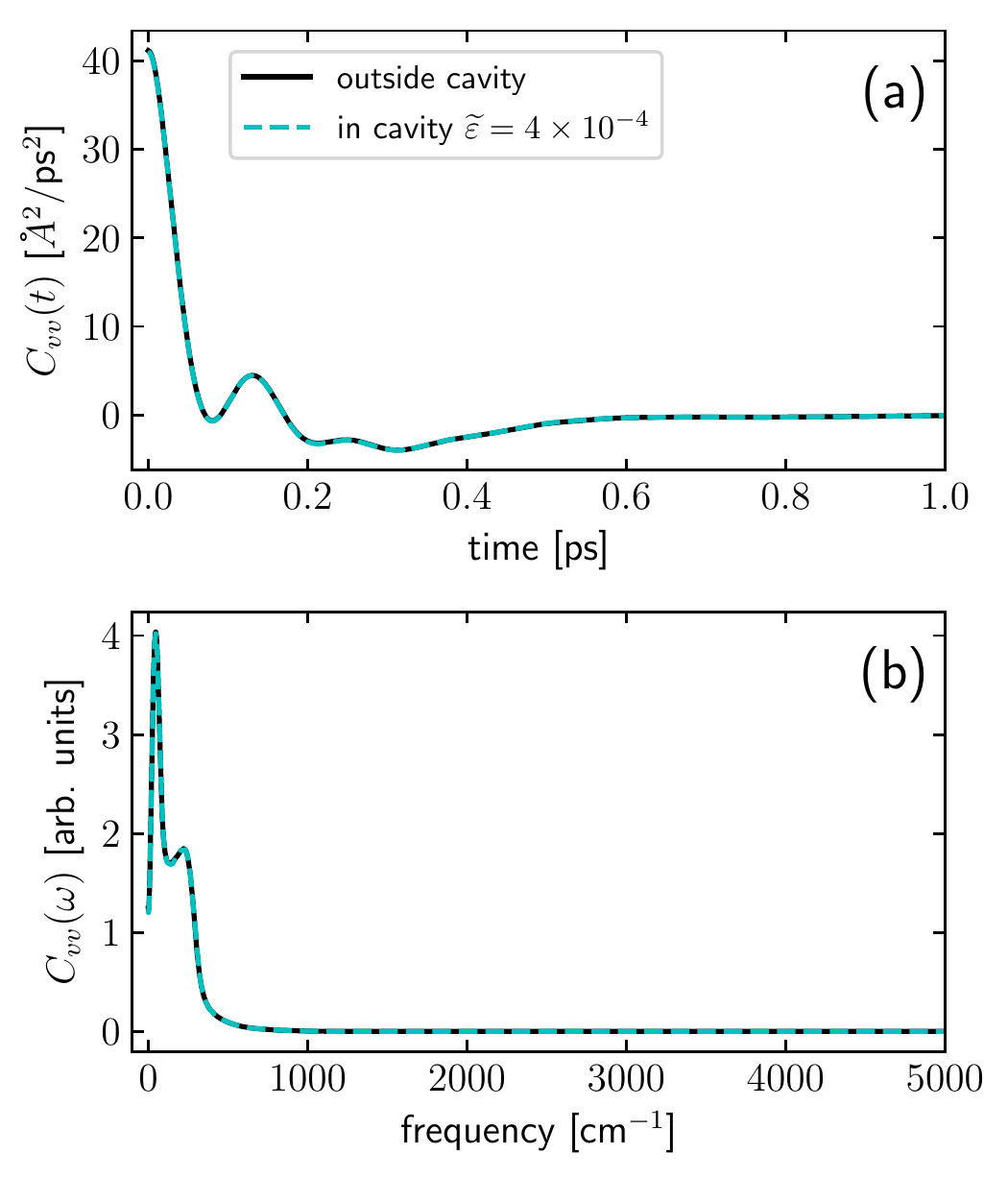}
		\caption{Velocity autocorrelation function (VACF) of the center of mass of individual \ch{H2O} molecules in the four-mode cavity plotted in the same manner as Fig. 6. All simulation details are the same as Fig. \ref{fig:IR_multimode}.}
		\label{fig:diffusion_multimode}
	\end{figure}

		\begin{figure}
		\centering
		\includegraphics[width=0.5\linewidth]{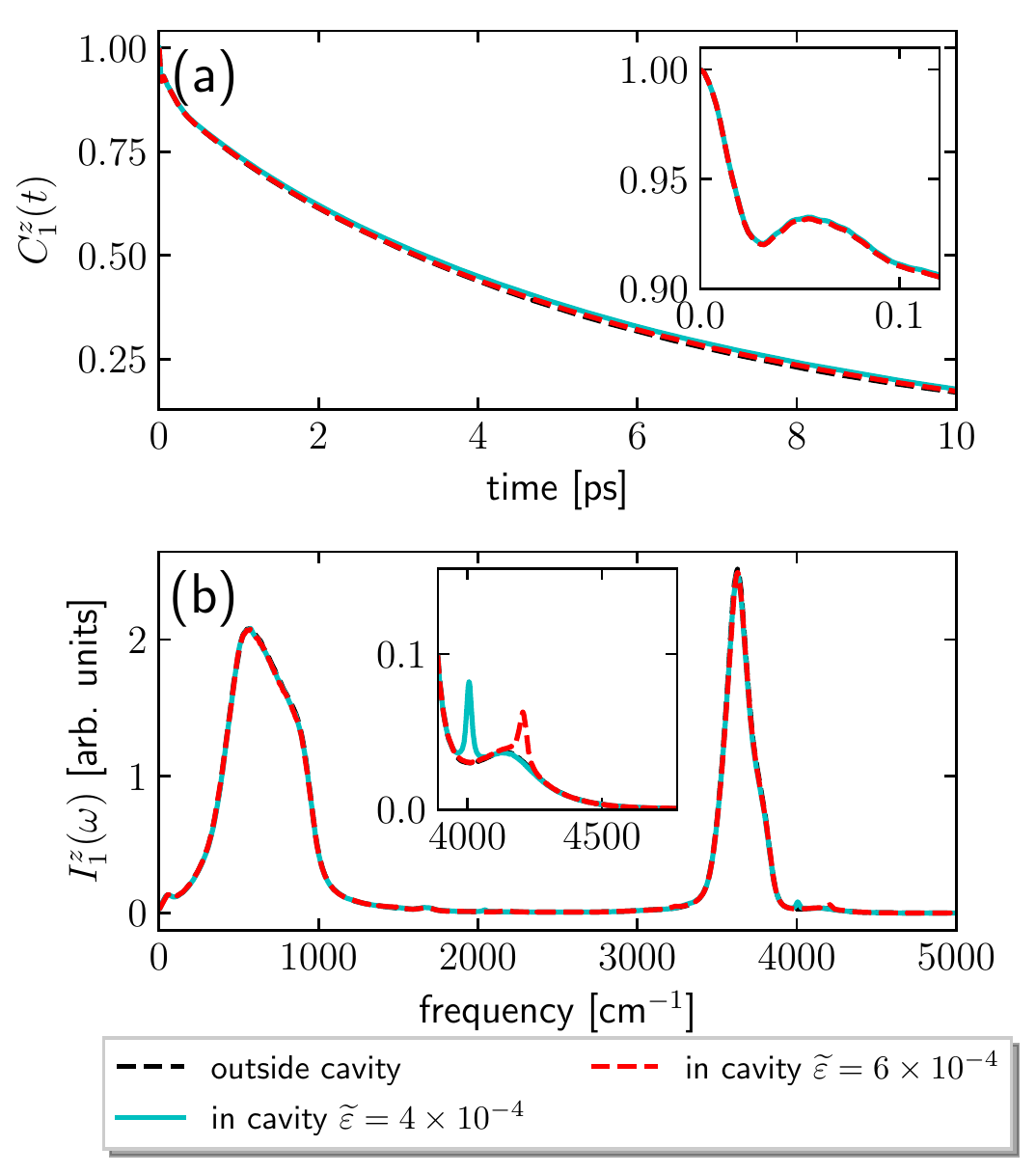}
		\caption{$z$-component of first-order orientational autocorrelation function (OACF) of individual \ch{H2O} molecules in the four-mode cavity plotted in the same manner as Fig. 7. All simulation details are the same as Fig. \ref{fig:IR_multimode}.}
		\label{fig:orientation_multimode}
	\end{figure}

	Finally, we report the VACF for the center-of-mass motion and the orientational autocorrelation function in Figs. \ref{fig:diffusion_multimode} and \ref{fig:orientation_multimode} in the same manner as Figs. 6 and 7. Clearly,  for a multimode cavity, the center-of-mass VACF is not changed, but the orientational autocorrelation function is slightly modified, similar to what was reported in Fig. 7 of the main text for the single-mode case.

	\newpage
	
	
	%